\documentclass[12pt]{article}
\usepackage{times}
\usepackage[margin=1in]{geometry}
\usepackage{indentfirst}
\usepackage{hyperref}
\usepackage{graphicx}
\usepackage{pdfpages}
\usepackage{tabularx}
\usepackage{booktabs}
\usepackage{rotating}
\usepackage{enumitem}
\usepackage{tikz}
\usetikzlibrary{arrows,shapes.arrows,positioning,shapes,decorations.text}
\usepackage{array}
\usepackage{soul}
\usepackage[english, cleanlook]{isodate}
\usepackage{setspace}
\usepackage[round]{natbib}
\def\checkmark{\tikz\fill[olive, scale=0.6](-.125,.375) -- (.25,0) -- (1.125,.875) -- (1, 1) -- (.25,.25) -- (0,.5) -- cycle;} 
\def\circlecheckmark{
\begin{tikzpicture}[scale = .6]
\fill[olive](-.125,.375) -- (.25,0) -- (1.125,.875) -- (1, 1) -- (.25,.25) -- (0,.5) -- cycle;
\draw[olive, line width = .1cm] (0.5,0.5) ellipse (0.9 and 0.9);
\end{tikzpicture}
} 
 
\newcolumntype{V}{>{\centering\arraybackslash} m{.12\linewidth}}

\newcolumntype{B}[1]{>{\centering\arraybackslash}b{#1}}

\title{Privacy, ethics, and data access: \\ A case study of the Fragile Families Challenge\footnote{Manuscript submitted to a special issue of \emph{Socius} about the Fragile Families Challenge. We thank the Board of Advisors of the Fragile Families Challenge for supporting the project and offering feedback throughout. Research reported in this publication was supported by the Russell Sage Foundation and the Eunice Kennedy Shriver National Institute of Child Health and Human Development of the National Institutes of Health under Award Number P2C-HD047879, and the National Science Foundation (NSF IIS Award 1704444 to AN). Funding for the Fragile Families and Child Wellbeing Study was provided by the Eunice Kennedy Shriver National Institute of Child Health and Human Development through grants R01-HD36916, R01-HD39135, and R01-HD40421 and by a consortium of private foundations, including the Robert Wood Johnson Foundation. We also thank Brandon Stewart, Alex Guerrero, Andrew Ledford, Julien Teitler, Sam Salganik, Amanda Slater, Prateek Mittal, and Vitaly Shmatikov for helpful conversations, and we thank Kristin Catena, Kate Jaeger, Ryan Vinh, Nathan Matias, and Jessica West for helpful feedback on drafts of this manuscript. Participants in the Princeton Sociology Proseminar, the Northeast Privacy Scholars Workshop, and the Center for Information Technology Policy reading group provided valuable feedback as well. This research was approved by the Institutional Review Board of Princeton University (Protocol 8061). The content of this paper is solely the responsibility of the authors and does not necessarily represent the views of anyone else.}}
\author{Ian Lundberg\textsuperscript{a}, Arvind Narayanan\textsuperscript{b}, Karen Levy\textsuperscript{c}, and Matthew J. Salganik\textsuperscript{a}}
\date{
\begin{footnotesize}
\textsuperscript{a}Department of Sociology, Princeton University \\
\textsuperscript{b}Department of Computer Science, Princeton University \\
\textsuperscript{c}Department of Information Science, Cornell University \\
Direct correspondence to \href{mailto:ilundberg@princeton.edu}{ilundberg@princeton.edu} \end{footnotesize}\\\vspace{1cm} Last updated: \today\vspace{.5cm}}

\begin{document}
\begin{singlespacing}

\maketitle
\newpage
\tableofcontents
\end{singlespacing}

\newpage
\section*{Abstract}
Stewards of social science data face a fundamental tension. On one hand, they want to make their data accessible to as many researchers as possible to facilitate new discoveries.  At the same time, they want to restrict access to their data as much as possible in order to protect the people represented in the data.  In this paper, we provide a case study addressing this common tension in an uncommon setting: the Fragile Families Challenge, a scientific mass collaboration designed to yield insights that could improve the lives of disadvantaged children in the United States. We describe our process of threat modeling, threat mitigation, and third-party guidance.  We also describe the ethical principles that formed the basis of our process.  We are open about our process and the trade-offs that we made in the hopes that others can improve on what we have done.

\section{Introduction}

\begin{figure}
\centering
\includegraphics[width=.6\textwidth]{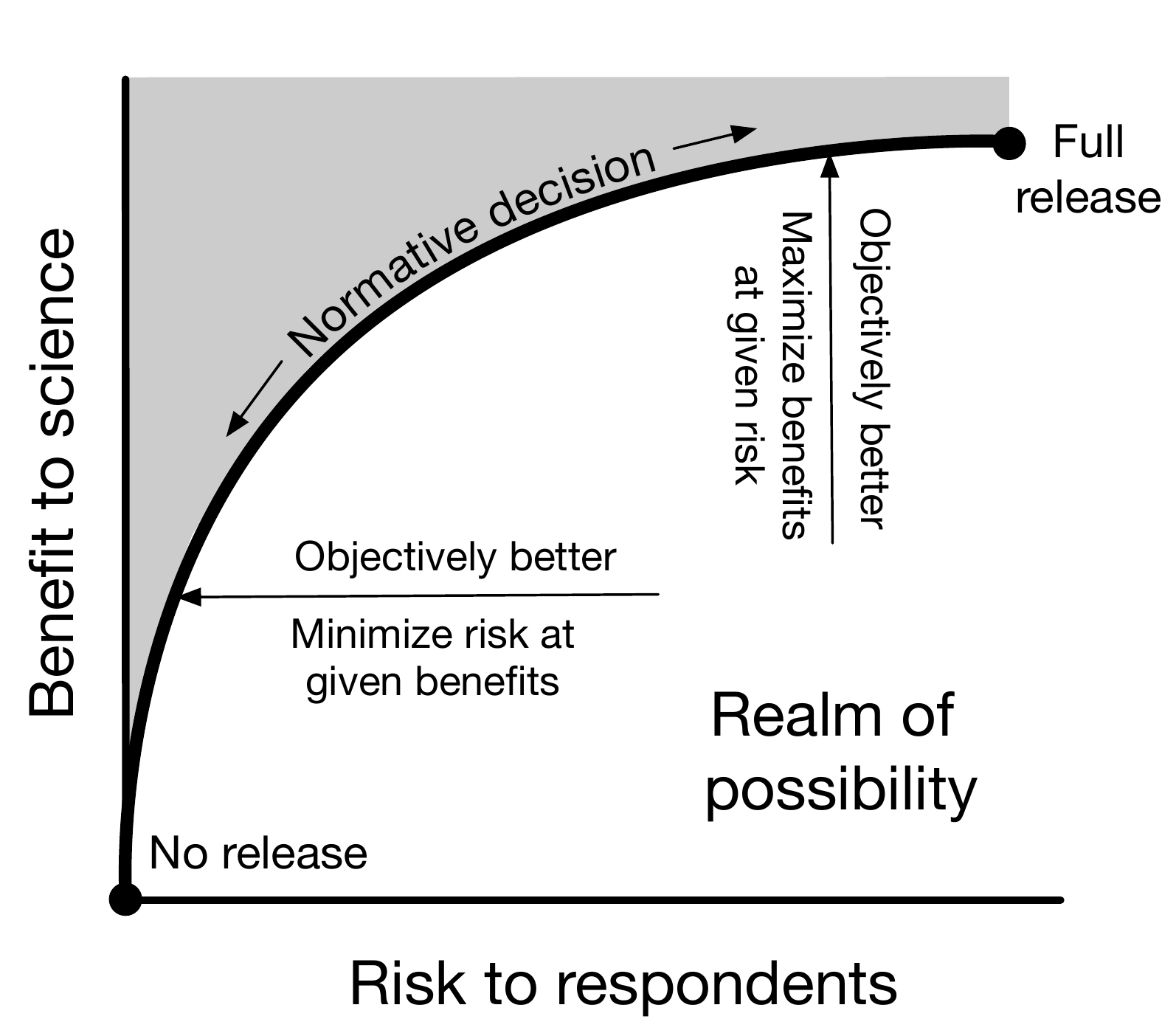}
\caption{\textbf{Data access involves a fundamental tension between risk to respondents and benefits to science.}
If research requires releasing data, then those who manage the data face a tension between risk to respondents and benefits to science.  It is objectively best to maximize benefits at a given level of risk and objectively best to minimize risk at a given level of benefits. On the frontier, the balance between risk and benefits becomes a normative question. The frontier is curved because we expect that, at low levels of risk, slight increases in risk might yield especially large benefits. For instance, moving from no release of data to release of a highly redacted form of the data might yield substantial benefits. At higher levels of risk, the returns to increased risk may be smaller. For instance, including respondents' addresses in the data release would substantially increase risk with only minimal benefits.  We emphasize that this curve is merely a heuristic device; in realistic situations it is difficult---perhaps impossible---to define and quantify risks and benefits \citep{lambert1993,karr2006,cox2011,skinner2012,goroff2015,narayanan2016}.  Further, many researchers are developing techniques, such as differential privacy \citep{dwork2008}, that try to improve the trade-offs.}
\label{fig:tensionFig}
\end{figure}

Social data---data about people---can be both valuable and dangerous.  On one hand, they can be used to advance scientific understanding and yield insights that can benefit society.  On the other hand, they can be used in ways that violate privacy and lead to other harms.  Stewards of social data, therefore, face a fundamental tension.  At one extreme, a data steward could share a complete dataset publicly with everyone.  This \emph{full release} approach maximizes the potential for scientific discovery, but it also maximizes risk to the people whose information is in the dataset.  At the other extreme, a data steward could share the data with no one.  This \emph{no release} approach minimizes risk to participants, but it also eliminates benefits that could come from the responsible use of the data.  In between these two extremes---no release and full release---there are a variety of intermediate solutions, which involve balancing risk to participants and benefits to science (Figure \ref{fig:tensionFig}).  In this paper, we present a case study describing how we balanced this trade-off between risks and benefits when we served as data steward during the Fragile Families Challenge.  We hope that this case study will benefit a variety of data stewards, including those within universities, companies, and governments.  We also hope that this case study will benefit policymakers who seek to enable responsible data access and researchers who seek responsible access to detailed and potentially sensitive social data.

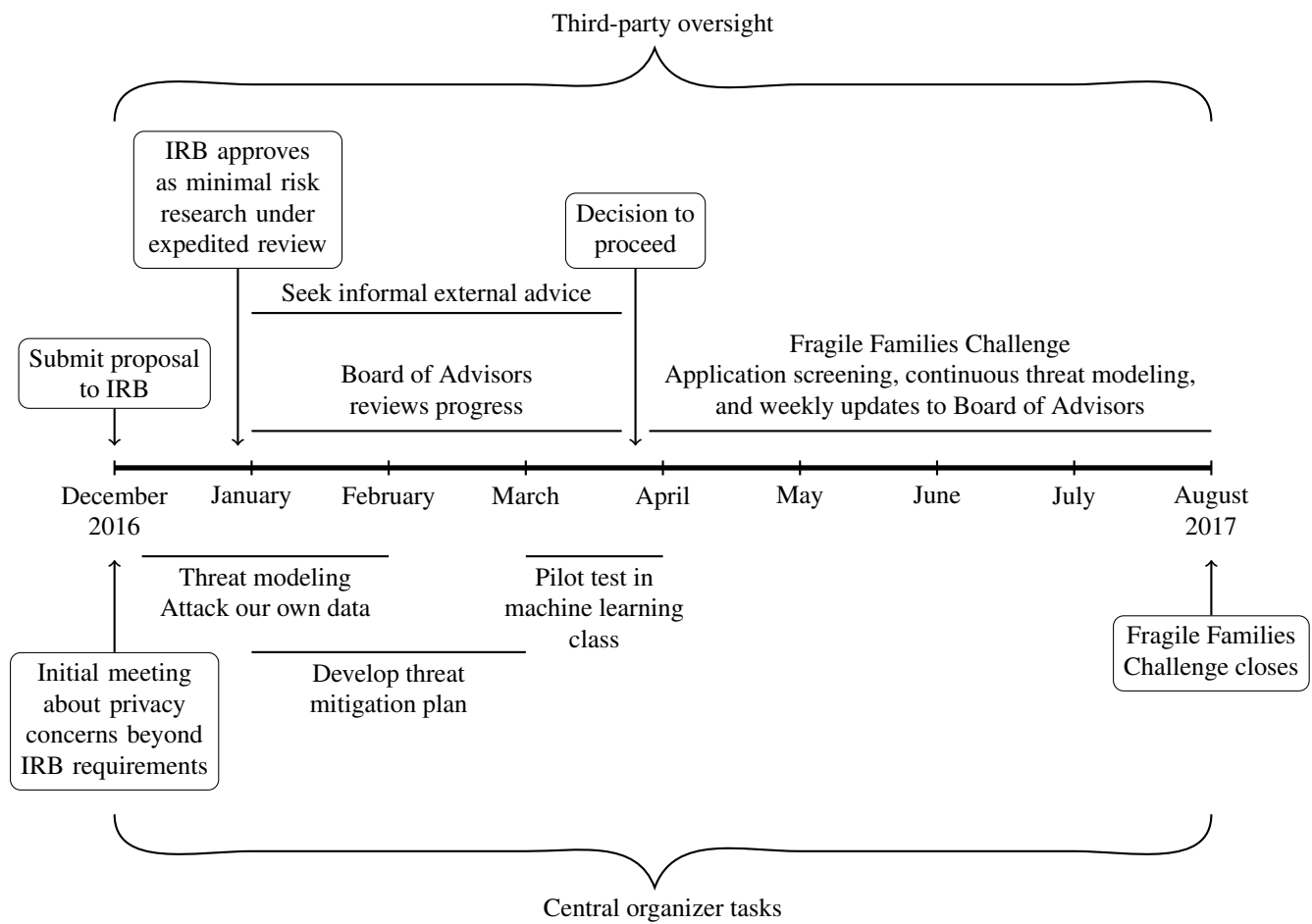
\begin{figure}
\centering
\begin{tikzpicture}[x = .9\textwidth]
\draw[line width = 2pt] (0,0) -- (1,0);
\draw[thick] (0,-.1) -- (0,.1);
\draw[thick] (1/8,-.1) -- (1/8,.1);
\draw[thick] (2/8,-.1) -- (2/8,.1);
\draw[thick] (3/8,-.1) -- (3/8,.1);
\draw[thick] (4/8,-.1) -- (4/8,.1);
\draw[thick] (5/8,-.1) -- (5/8,.1);
\draw[thick] (6/8,-.1) -- (6/8,.1);
\draw[thick] (7/8,-.1) -- (7/8,.1);
\draw[thick] (8/8,-.1) -- (8/8,.1);
\node[anchor = north, align = center, font = \footnotesize] at (0,-.1) {December\\2016};
\node[anchor = north, align = center, font = \footnotesize] at (1/8,-.1) {January};
\node[anchor = north, align = center, font = \footnotesize] at (2/8,-.1) {February};
\node[anchor = north, align = center, font = \footnotesize] at (3/8,-.1) {March};
\node[anchor = north, align = center, font = \footnotesize] at (4/8,-.1) {April};
\node[anchor = north, align = center, font = \footnotesize] at (5/8,-.1) {May};
\node[anchor = north, align = center, font = \footnotesize] at (6/8,-.1) {June};
\node[anchor = north, align = center, font = \footnotesize] at (7/8,-.1) {July};
\node[anchor = north, align = center, font = \footnotesize] at (8/8,-.1) {August\\2017};
\node[anchor = north, align = center, font = \footnotesize, draw, rounded corners] (closed) at (8/8,-2) {Fragile Families\\Challenge closes};
\draw[->, thick] (closed) -- (1,-1.25);
\node[font = \footnotesize] (oversight) at (4/8, 6) {Third-party oversight};
\draw[thick] (0,4.7) to[out = 90, in = 180] (1/8, 5.2) -- (3/8, 5.2) to[out = 0, in = -110] (oversight) to[out = -70, in = 180] (5/8,5.2) -- (7/8, 5.2) to[out = 0, in = 90] (1,4.7);
\node[anchor = south, align = center, font = \footnotesize, draw, rounded corners] (irb_application) at (0/8,.8) {Submit proposal\\to IRB};
\draw[->, thick] (irb_application) -- (0,.3);
\node[anchor = south, text width=1in, align = center, font = \footnotesize, draw, rounded corners] (irb_approval) at (0.9/8,2.7) {IRB approves as minimal risk research under expedited review};
\draw[->, thick] (irb_approval) -- (0.9/8,.3);
\draw[thick] (1/8,.5) -- (3.7/8, .5) node[midway, above, font = \footnotesize,align=center] {Board of Advisors\\reviews progress};
\node[anchor = south, draw, rounded corners, font = \footnotesize, align = center] (decision) at (3.8/8, 2.7) {Decision to\\proceed};
\draw[->, thick] (decision) -- (3.8/8,.3);
\draw[thick] (3.9/8,.5) -- (8/8, .5) node[midway, above, font = \footnotesize,align=center] {Fragile Families Challenge\\Application screening, continuous threat modeling,\\and weekly updates to Board of Advisors};
\draw[thick] (1/8,2.1) -- (3.7/8, 2.1) node[midway, above, font = \footnotesize,align=center] {Seek informal external advice};
\node[font = \footnotesize] (oversight) at (4/8, -6) {Central organizer tasks};
\draw[thick] (0,-4.7) to[out = -90, in = 180] (1/8, -5.2) -- (3/8, -5.2) to[out = 0, in = 110] (oversight) to[out = 70, in = 180] (5/8,-5.2) -- (7/8, -5.2) to[out = 0, in = -90] (1,-4.7);
\node[anchor = north, text width=1in, align = center, font = \footnotesize, draw, rounded corners] (initial_meeting) at (0,-2.5) {Initial meeting about privacy concerns beyond IRB requirements};
\draw[->, thick] (initial_meeting) -- (0,-1.25);
\draw[thick] (0.2/8,-1.2) -- (2/8, -1.2) node[midway, below, font = \footnotesize,align=center] {Threat modeling\\Attack our own data};
\draw[thick] (1/8,-2.5) -- (3/8, -2.5) node[midway, below, font = \footnotesize,align=center] {Develop threat\\mitigation plan};
\draw[thick] (3/8,-1.2) -- (4/8, -1.2) node[midway, below, font = \footnotesize,align=center] {Pilot test in\\machine learning\\class};
\end{tikzpicture}
\caption{\textbf{Timeline of the privacy and ethics process for the Fragile Families Challenge.} Boxed nodes represent events occurring at a specific point in time, such as the decision to proceed. Lines represent activities which occurred over the course of a period of time, such as seeking informal external advice.}
\label{fig:timeline}
\end{figure}

The Fragile Families Challenge is a scientific mass collaboration involving hundreds of researchers.  During this mass collaboration, a diverse group of social scientists and data scientists worked with a common dataset that contained detailed information about the lives of about 5,000 families in the United States, many of whom were disadvantaged.  The detailed nature of the data made the project particularly valuable for developing knowledge about the lives of disadvantaged families, yet these very features also heightened concerns about privacy and ethics. In other words, the Challenge brought the tension between risks and benefits into sharp focus.  In this paper, we provide no single solution to the fundamental tension between access and privacy; instead, we describe our process of addressing it. More specifically, this paper describes the privacy and ethics audit that we conducted from December 2016 through March 2017, as well as steps we carried out during the Challenge from March through August 2017 (see timeline in Figure \ref{fig:timeline}). 

\begin{figure}
\centering
\begin{tikzpicture}[x = 1in, y = .7in]
\node (threatModeling) at (0,1) {Threat modeling};
\node[align=center] (threatMitigation) at (1,0) {Threat\\mitigation};
\node[align=center] (third) at (-1,0) {Third-party\\guidance};
\draw[<->,line width = 2pt] (threatModeling) -- (threatMitigation);
\draw[<->,line width = 2pt] (third) -- (threatMitigation);
\draw[<->,line width = 2pt] (third) -- (threatModeling);
\draw[line width = 2pt] (0,.4) ellipse (1.75in and .8in);
\node[fill=white] at (0, -.65) {Ethical framework};
\end{tikzpicture}
\caption{\textbf{Privacy and ethics audit for the Fragile Families Challenge.} Our process involved (a) threat modeling to make precise our fear of re-identification, (b) threat mitigation to reduce risk, and (c) the guidance of third parties such as the IRB at Princeton University and the Fragile Families Challenge Board of Advisors. The entire process was undertaken within an ethical framework which we describe in Section \ref{ethics}. While the paper is written linearly, we emphasize that these steps were not taken linearly; we cycled through all the steps many times.}
\label{fig:process}
\end{figure}
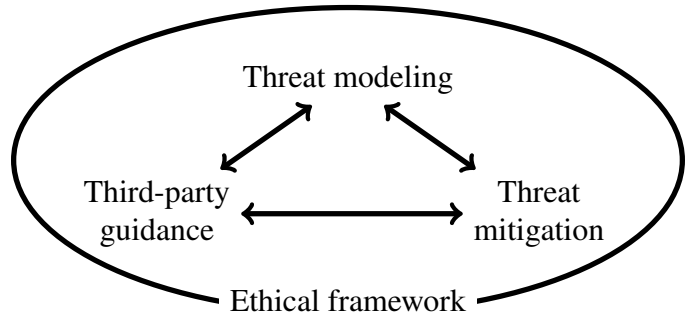

The paper is divided into 8 sections.  Section \ref{background} provides more background about the data and the Challenge.  Section \ref{threat_modeling_and_threat_mitigation} describes our threat modeling and threat mitigation strategies, and Section \ref{responsePlan} describes our response plan in case our mitigations were ineffective. Section \ref{guidance} summarizes our mechanisms for third-party guidance.  Section \ref{ethics} discusses the ethical principles that guided our thinking. Section \ref{decision} describes our ultimate decision to conduct the Challenge.  Section \ref{conclusion} concludes.  Although the paper is written linearly, the real process---summarized in Figure \ref{fig:process}---cycled through all of these steps many times.  We are open about our process and the trade-offs that we made in the hope that others can improve on what we have done.

\section{Background}
\label{background}

\subsection{Fragile Families Study}
\label{data}

The Fragile Families Challenge builds on the Fragile Families and Child Wellbeing Study (hereafter Fragile Families Study), a longitudinal study of 4,898 families.  The study began with a probability sample of newborns in 20 large U.S. cities, of which 16 cities form a probability sample of all U.S. cities with populations over 200,000 \citep{reichman2001}.  For more than 15 years, researchers have followed these families to collect information related to child and family development as reported by the child as well as the child's mother, father, primary caregiver, and teachers.  These rich longitudinal data have already been used in hundreds of published papers and dozens of dissertations on aspects of urban poverty including multiple-partner fertility \citep{carlson2006}, multigenerational households \citep{pilkauskas2012}, paternal incarceration \citep{wildeman2009}, housing instability \citep{desmond2015}, and neighborhood disadvantage \citep{donnelly2017}.\footnote{For a complete list of research using the Fragile Families Study, see \url{https://ffpubs.princeton.edu/}.}

Four features of the Fragile Families Study were particularly relevant to the design and conduct of the Fragile Families Challenge.  Many of these features are common in large-scale social science datasets, but may not be common in datasets held by companies and governments.  These features also make these data different from some types of data commonly considered in privacy research.

First, these data were collected with informed consent.  Parents explicitly agreed to join the study and made this agreement on behalf of their children.  Further, the children themselves provided their assent to participate once they were old enough. These procedures were overseen by the Institutional Review Board of Princeton University.  Informed consent makes the Fragile Families Study different from many other cases privacy scholars have considered. For instance, a main critique of the use of Facebook data for research purposes has been the lack of informed consent; participants may not expect their activity on Facebook to be used in research \citep{zimmer2010}.  In this case, however, respondents learned about the goals of the study and gave explicit permission for the information they provided to be used by researchers.

\begin{sidewaysfigure}
\centering
\begin{tikzpicture}[x = .5in, y = .6in, style={font=\footnotesize}]
\draw (-13, 0) rectangle (-11,2.5);
\node[align=center] at (-12, 2.8) {\textbf{Raw Data}};
\node[draw=black] at (-12.5, 1.955) {\scriptsize Birth};
\node[draw=black] at (-11.5, 1.955) {\scriptsize Age 1};
\node[draw=black] at (-12.5, 1.35) {\scriptsize Age 3};
\node[draw=black] at (-11.5, 1.35) {\scriptsize Age 5};
\node[draw=black] at (-12.5, 0.625) {\scriptsize Age 9};
\node[draw=black] at (-11.5, 0.625) {\scriptsize Age 15};
\node[align=center] at (-12, -.25) {\textbf{Identifiers}};
\node[draw, align=center, font = \scriptsize, minimum width = 1in] at (-12, -.75) {Stored separately\\and never shared};
\draw[thick, ->] (-10.8,2.4) -- (-8.95,2.4);
\node[text width = 1in,anchor = north] at (-9.75,2.5) {\begin{itemize}[leftmargin=*]
\item Add noise to geocode data
\item Split by domain
\end{itemize}};
\draw[dashed] (-8.5, -1) rectangle (-1.5, 3.1);
\node[align=center] at (-5, -.5) {Data provided to researchers\\before the Fragile Families Challenge};
\draw (-8, 0) rectangle (-6,2.5);
\node[draw=black] at (-7, 1.975) {\scriptsize Genetic};
\node[draw=black] at (-7, 1.35) {\scriptsize Geocode};
\node[draw=black] at (-7, .625) {\scriptsize All other};
\node[align=center] at (-7, 2.8) {\textbf{Restricted Files}};
\draw[thick, ->] (-5.8,2.4) -- (-4.2,2.4);
\node[text width = .8in,anchor = north] at (-5,2.5) {\begin{itemize}[leftmargin=*]
\item Redact obvious indirect identifiers
\item Change pseudo-ID
\end{itemize}};
\node[align=center] at (-3, 2.8) {\textbf{Basic Files}};
\draw (-4, 0) rectangle (-2, 2.5);
\node[draw=black] at (-3.5, 1.955) {\scriptsize Birth};
\node[draw=black] at (-2.5, 1.955) {\scriptsize Age 1};
\node[draw=black] at (-3.5, 1.35) {\scriptsize Age 3};
\node[draw=black] at (-2.5, 1.35) {\scriptsize Age 5};
\node[draw=black] at (-3, 0.625) {\scriptsize Age 9};
\draw[thick, ->] (-1.05,2.4) -- (.8,2.4);
\node[text width = 1in,anchor = north] at (-.25,2.5) {\begin{itemize}[leftmargin=*]
\item Redact
\item Coarsen 
\item Add noise
\item Change pseudo-ID
\item Add 6 age 15 outcomes
\end{itemize}};
\node[align=center] at (2, 2.8) {\textbf{Challenge Files}};
\draw (1, 0) rectangle (3, 2.5);
\node[draw=black,align=center] at (2, 1.8) {\scriptsize Background\\\scriptsize predictors\\ \scriptsize(birth to age 9)};
\node[draw=black,align=center] at (2, 0.7) {\scriptsize Training\\\scriptsize outcomes\\\scriptsize(age 15)};
\end{tikzpicture}
\caption{\textbf{Versions of the Fragile Families and Study data.} There are several versions of the data. These versions are made available to researchers depending on their particular needs. The Raw Data are used only by survey administrators, and are stored in separate files to reduce the risk of a breach. For instance, no data file contains respondent names and survey responses. All files given to researchers have names and other obvious identifiers removed and noise added to any data indicating place of residence. Among the files available to researchers, the Restricted Files provide the most information but are hardest to access. Researchers obtain a Restricted File through an intensive application and screening process. After this process, researchers are only given the part of the Restricted Files needed for their particular project. Most researchers' projects can be completed using only the Basic Files, for which one still must apply by proposing a research project. To create the Challenge Files we made modifications to the Basic Files.}
\label{fig:versions}
\end{sidewaysfigure}
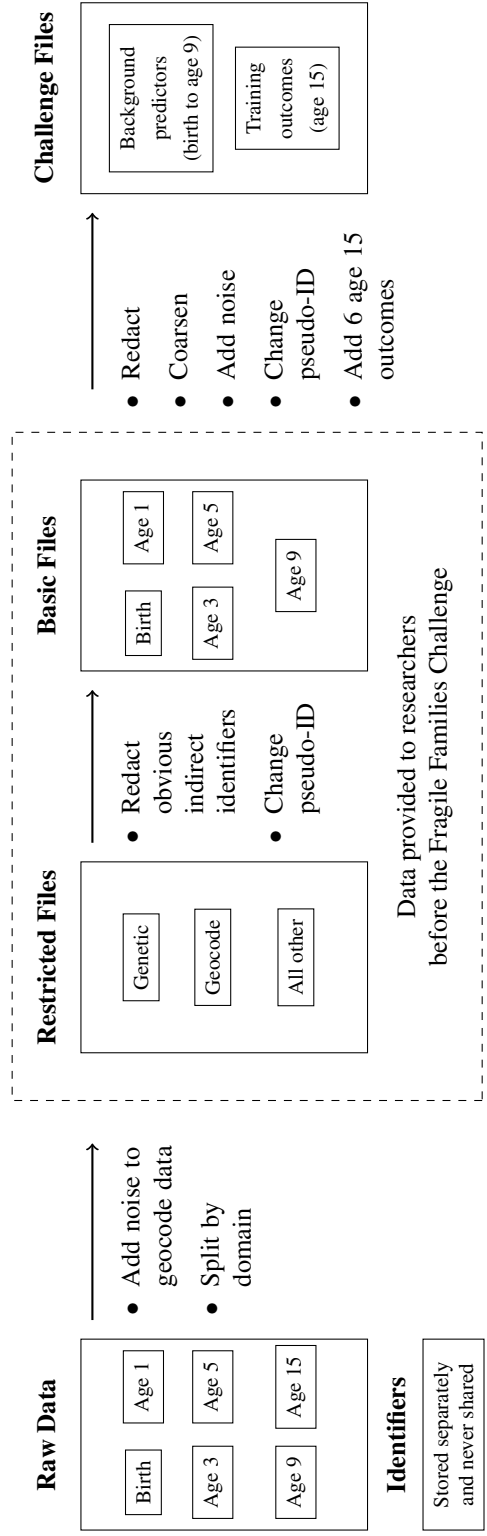

Second, these data are already available to researchers through an established system.  This data access system, which is overseen by the Institutional Review Board of Princeton University, has already been used by thousands of researchers for more than 15 years.  The current data access system follows a tiered model in which there are Basic Files and Restricted Files (Figure \ref{fig:versions}).  The two main differences between the Basic and Restricted Files are the application process and the types of data that are provided; in all cases, the data are stripped of obviously personally identifying information.\footnote{The Restricted Files may contain detailed geographic, genetic, or other data deemed especially sensitive or identifiable. Access to these files requires a detailed restricted data contract and a carefully vetted research proposal. The Fragile Families staff grants approval only to projects with research merit that can only be achieved with the restricted files. Data are shared only after researchers sign a data protection agreement, show that they have completed NIH-approved Protecting Human Research Participants training, provide evidence of approval from the Institutional Review Board of their institution, and detail a data protection plan summarizing how data will be used. Student applicants must apply with a faculty mentor who bears responsibility for violations of the agreement. Researchers approved in this process are given access to a tailored version of the data with detailed information only on the domains relevant to their research. For instance, a researcher studying neighborhood effects would be given neighborhood information but not genetic information, while a researcher studying epigenetics would be given genetic information but not neighborhood information; only researchers with projects involving both fields would be given data covering both domains. Fragile Families staff work with researchers to determine the particular variables needed for any given study.  The Basic Files exclude obvious personally identifying information and obvious geographic information, such as city of birth and residential location. To obtain these files, researchers must agree to a set of terms and conditions and propose a viable project which Fragile Families staff approve as potentially important social science research. Several thousand researchers have completed these procedure and were already using the data before the Challenge began.}  This system of tiered access served as our baseline as we designed and implemented the Fragile Families Challenge. We think it is reasonable to accept the current system as our baseline because this system developed over many years in the full view of the scientific community.

\begin{figure}
\centering\includegraphics[width = .8\textwidth]{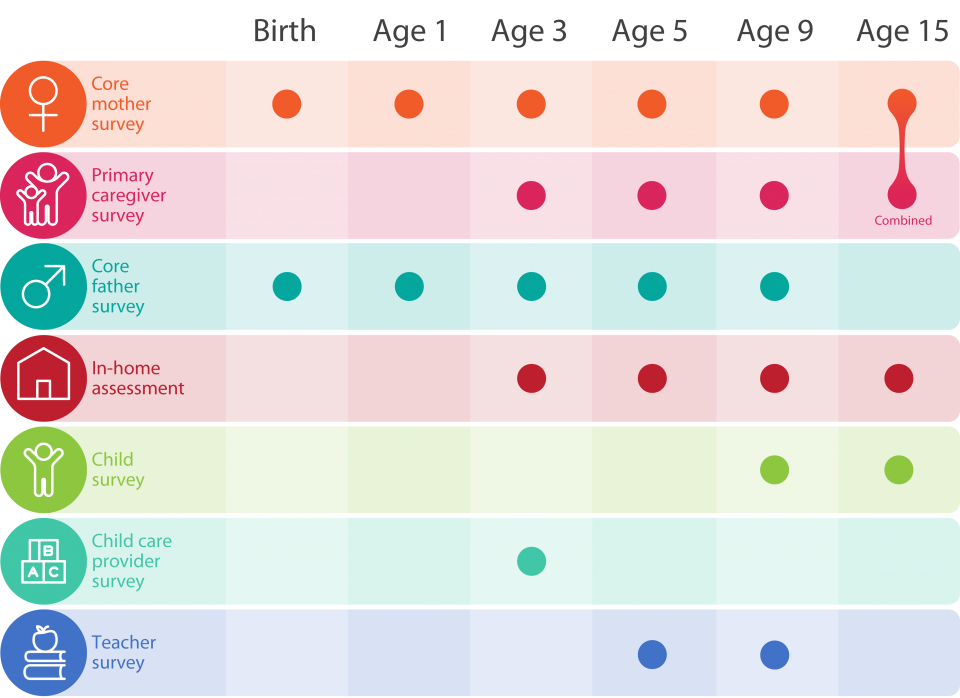}
\caption{\textbf{Domains covered in the Fragile Families Study basic files.} The number of substantive domains included makes the Fragile Families Study especially useful to social scientists. The number of domains also (a) facilitates re-identification because many possible auxiliary datasets may be used by an adversary and (b) increases the risk of harm in the event of re-identification because substantial information about respondents' private lives could become public. Source: \url{http://fragilefamilies.princeton.edu/documentation}}
\label{fig:domains}
\end{figure}

The third feature of the Fragile Families Study that shaped our design of the Challenge is that these data contain information about many people around the focal child---such as the mother, father, primary caregiver, and teacher---and contain information about many domains of the respondents' lives.  For example, the study collects information about the home environment, the school environment, teacher characteristics, parental criminal history, and child health, to name just a few domains (Figure \ref{fig:domains}).  The multi-domain, linked nature of the data increases its scientific value, but it also increases risks for two main reasons.  First, it creates many possible entry points for a re-identification attack (this risk will be described in more detail in Section \ref{worry}).  Second, the multi-domain nature of the data increases the harm that could occur if a re-identification attack was successful, because many potentially sensitive pieces of information would be revealed.  The linked, multi-domain nature of the data differs from the cases normally considered by privacy researchers, which usually involve a collection of individuals with information from a single domain (e.g., medical records).

The fourth and final feature that is relevant to the Fragile Families Challenge is that these data are frequently used in scientific and policy debates. For example, a recent National Academies of Sciences report on the effects of parental incarceration on children drew heavily on the Fragile Families Study \citep{travis2014}.  
Although these data have already been used extensively by social scientists, we thought that it would be possible to learn even more if a larger, more diverse group of researchers approached the data in a very different way.  This goal of increased scientific and policy impact was one of the main motivations for the Fragile Families Challenge.

\subsection{Fragile Families Challenge}
\label{FFC}

The Fragile Families Challenge is a mass collaboration that combines predictive modeling, causal inference, and in-depth interviews to yield insights that can improve the lives of disadvantaged children in the United States.  This paper---and the special issue of the journal \emph{Socius} in which this paper will be published---describes the first stage of the Fragile Families Challenge, which focuses on predictive modeling.  This predictive modeling stage follows an approach called the Common Task Framework which is used frequently in computer science \citep{donoho2017} and biomedical research \citep{saez-rodriguez2016}.  The Common Task Framework is a process that invites many researchers to participate in a unified task characterized by three key aspects: 1) a common predictive modeling goal using 2) a single dataset made available to all with 3) a well-defined scoring metric to evaluate contributions. This process often yields better predictive performance than any individual researcher can realize working alone (e.g., \citealt{bennett2007}) and often leads to new scientific and methodological insights (e.g., \citealt{feuerverger2012}).  \citet{donoho2017} went so far as to describe the Common Task Framework as part of ``the `secret sauce' of machine learning.'' 

\begin{figure}
\centering\includegraphics[width = .8\textwidth]{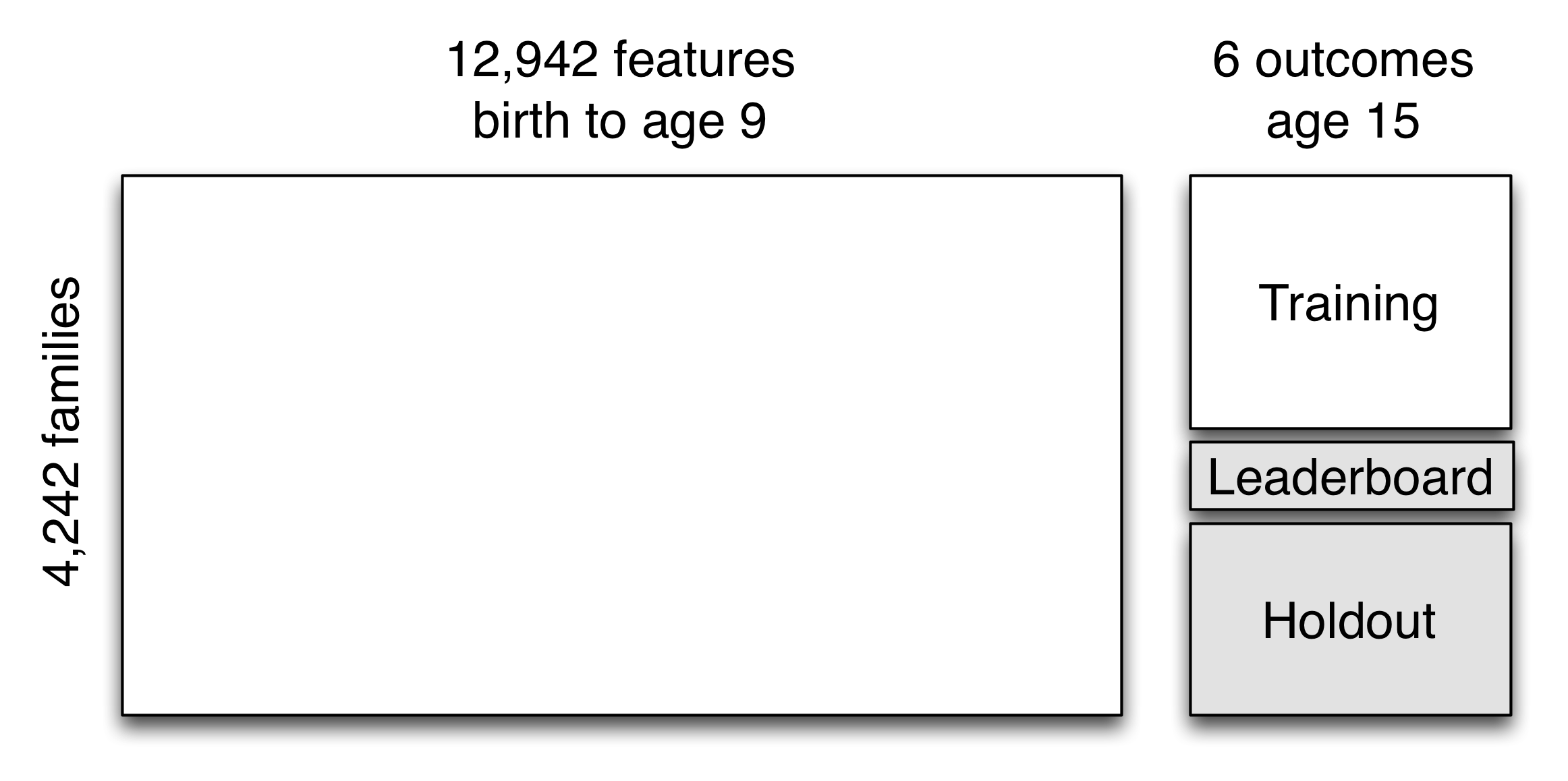}
\caption{\textbf{Fragile Families Challenge data structure.} Participants built models predicting the age 15 outcomes using data collected while the focal child was age 9 and younger. We provided participants with the data represented by the white boxes. Submissions were scored based on their predictive performance (mean squared error) for the observations represented by the gray boxes, which were available only to organizers. The leaderboard set contained 1/8 of all observations and was used to provide instant feedback on submissions. The holdout set contained 3/8 of observations and was used to produce final scores for all submitted models at the end of the Challenge.}
\label{fig:ffc_design_matrix_ml}
\end{figure}

As described in more detail in the introduction to the special issue in which this paper will be published \citep{salganik2018intro}, we set out the goal of using the data collected from a family at the birth of the child up to when the child was age 9 to predict data from the family when the child was 15 years old.  This age 15 data had been collected but was not yet available to participants (Figure \ref{fig:ffc_design_matrix_ml}).  The existence of this collected but otherwise unavailable data is critical for the Common Task Framework, and fortunately all longitudinal social surveys present this possibility every time a new wave of data has been collected.  Among the many possible outcomes in the year 15 data, we asked participants in the Challenge to predict six key outcomes: grade point average (GPA) of the child, grit of the child, material hardship of the family, whether the family was evicted from their home, whether the primary caregiver participated in job training, and whether the primary caregiver lost his or her job.  The choice of these outcomes was driven by ethical considerations and our scientific goals, and each outcome is described in more detail in the introduction to the special issue in which this paper will be published.

As is typical in projects using the Common Task Framework, we split the year 15 data for these six outcomes into three groups: 1) a training set which we provided to participants, 2) a leaderboard set which participants could access during the Challenge, and 3) a holdout set which participants could not access until the first stage of the Challenge was complete \citep{hardt2015} (Figure \ref{fig:ffc_design_matrix_ml}).  Participants in the Challenge received the training set and a specially constructed background data file that contained information about the family from birth to age 9 (more on the construction of this file is described in Section \ref{modifications}).   This background file included 4,242 families and 12,942 variables about each family.  The high-dimensional nature of the data---more predictors than observations---is a result of the linked, multiple-domain nature of the data (as discussed above) and has important implications for privacy (as discussed below).  

Challenge participants used the background data file and training data to build statistical or machine learning models.  They used these models to predict the holdout data (e.g., grade point average at age 15).  We measured the quality of these predictions using mean squared error.\footnote{The mean squared error is a common scoring metric, and it can be written as $\frac{1}{n} \sum_{i=1}^n (y_i - \hat{y}_i)^2$  where where $\hat{y}_i$ is the predicted value for person $i$, $y_i$ is the true value for person $i$, and $n$ is the number of people in the holdout set.  For binary outcomes, mean squared error is sometimes called Brier score \citep{brier1950}.}  

The immediate goal of this stage of the Fragile Families Challenge was to find the most accurate predictive model for the six outcomes. Given the nature and size of the data---thousands rather than millions of observations---we wanted to learn the extent to which machine learning methods would improve predictive performance beyond the generalized linear models typically applied by social scientists. This explicit focus on prediction is atypical for social science, but substantial current scholarship argues that prediction is important for both scientific and policy reasons \citep{breiman2001,shmueli2010,watts2014,kleinberg2015,kleinberg2017,mullainathan2017,hofman2017}.

In our case, the immediate goal of prediction was important to prepare for a long-term goal of explanation and hypothesis generation. As will be described in future papers, these predictions will be used to target qualitative, in-depth interviews with families who reported unexpected outcomes. We hope that these interviews will help us discover important and currently unmeasured factors and generate hypotheses about how these may impact the lives of disadvantaged families. We also hope that the interviews will inform the credibility of the assumptions required to draw causal inferences from survey data in a setting in which thousands of pretreatment variables are potentially involved in confounding.

\subsection{Why worry?}
\label{worry}

The data from the Fragile Families Study were collected with informed consent and are already provided to researchers under a well-established system, all of which has been overseen by the Institutional Review Board of Princeton University.  More generally, survey data of this type have been provided to researchers in a similar fashion for over 50 years.  Why should we worry about the risk of re-identification in the Fragile Families Challenge?

Quite simply: all data are potentially identifiable.  This possibility was made clear to us in one of our first meetings when a member of our team (Narayanan) proposed the following hypothetical scenario.  Imagine an all-powerful and evil business magnate like Lex Luthor, Superman's archenemy.  Further imagine that Lex Luthor heard about the Fragile Families Challenge and wished to re-identify the data. Lex could invest billions of dollars to conduct a census of every child born from 1998 through 2000 in the cities covered by the Fragile Families Study. Then Lex could merge his census with the Fragile Families Challenge data, identify everyone, and then learn everything about them in the Fragile Families Study.

This hypothetical attack from Lex Luthor illustrates two important points.  First, it illustrates a common pattern in re-identification attacks.  Data that have been \emph{de-identified}, meaning stripped of obviously identifying information in an effort to protect privacy, can often be \emph{re-identified} by linkage to an auxiliary dataset that contains identifying information.  Through this process of merging, the apparently anonymous data are re-identified \citep{ohm2010}.  This hypothetical attack also illustrates that the safety of a given dataset depends not just on that dataset but on all the auxiliary data that exist today and may exist in the future \citep{narayanan2010}. De-identification of a dataset does not guarantee anonymity.

\begin{figure}
A. Sweeney's \citeyear{sweeney2002} re-identification of Massachusetts medical records \vskip .2cm
\begin{center}
\begin{tikzpicture}[x = 1in, y = 1in]
\draw[blue, fill=blue, draw = black, fill opacity = .1, line width = 0pt] (0.5,0) ellipse (1in and 1in);
\draw[red, fill=red, draw = black, fill opacity = .1, line width = 0pt] (1.5,0) ellipse (1in and 1in);
\node[align=center] at (0,0) {Prescriptions\\Diagnoses};
\node[align=center] at (2,0) {Name\\Address};
\node[align=center] at (1,0) {Date of birth\\Zip code\\Sex};
\node[align=center,fill = blue, fill opacity = .1, text opacity = 1, anchor=south] at (0.4,1.1) {Medical records};
\node[align=center,fill = red, fill opacity = .1, text opacity = 1, anchor=south] at (1.6,1.1) {Voting records};
\end{tikzpicture}
\end{center}
B. Malin and Sweeney's \citeyear{malin2004} re-identification of genomics data
\begin{center}
\begin{tikzpicture}[x = 1.2in, y = 1in]
\draw[blue, fill=blue, draw = black, fill opacity = .1, line width = 0pt] (0.5,0) ellipse (1.1in and 1in);
\draw[yellow, fill=green, draw = black, fill opacity = .2, line width = 0pt] (1.5,0) ellipse (1.1in and 1in);
\draw[red, fill=red, draw = black, fill opacity = .1, line width = 0pt] (2.5,0) ellipse (1.1in and 1in);
\node[align=center] at (0,0) {DNA};
\node[align=center] at (1,0) {Collection\\time and\\place};
\node[align=center] at (2,0) {Date of birth\\Zip code\\Sex};
\node[align=center] at (3,0) {Name\\Address};
\node[align=center,black,fill=blue,fill opacity = .1, text opacity = 1, anchor=south] at (0.4,1.1) {DNA records};
\node[align=center,fill=green, fill opacity = .2, text opacity = 1, anchor=south] at (1.5,1.1) {Hospital\\discharge data};
\node[align=center,fill = red, fill opacity = .1, text opacity = 1, anchor=south] at (2.6,1.1) {Voting records};
\end{tikzpicture}
\end{center}
\caption{\textbf{Re-identification examples with de-identified data.} In both examples, the adversary succeeded because key variables were available in both the de-identified dataset (blue) and an identified auxiliary dataset (red).  In addition to merges between two datasets \citep{sweeney2002}, re-identification can proceed through a chain of auxiliary datasets \citep{malin2004}.}
\label{fig:sweeney}
\end{figure}
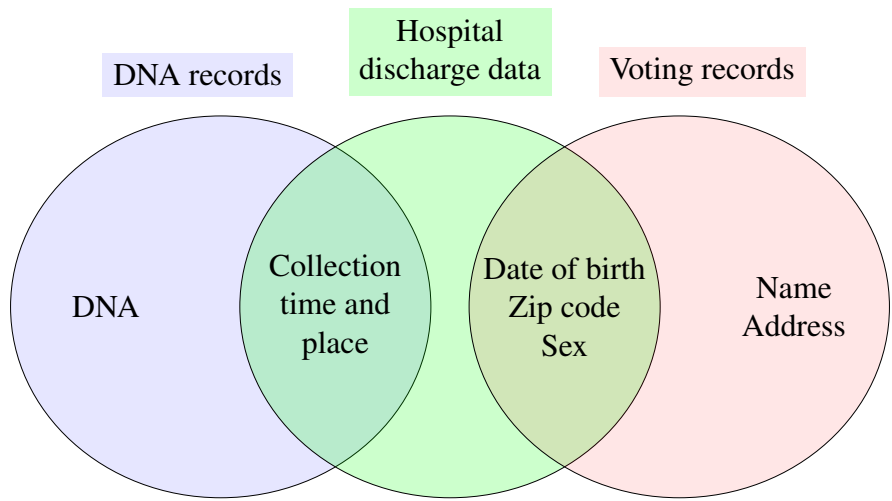

Re-identification attacks like the one performed by Lex Luthor are not merely hypothetical risks, unfortunately.  Although we do not know the frequency with which these attacks occur ``in the wild,'' we do know that academic privacy researchers have conducted and published similar attacks (we will discuss their motivations for these attacks in Section \ref{privacyresearcher}).\footnote{For an attempt to estimate the rate of re-identification attacks that have been published in the academic literature, see \citet{elemam2011}.}  Two prominent examples come from the research of Latanya Sweeney.  First, while a graduate student at MIT, \citet{sweeney2002} was able to re-identify apparently anonymous medical records that were provided to researchers by the Massachusetts Group Insurance Commission.  She did this by combining the apparently anonymous medical records, which contained date of birth, zip code, and sex, with non-anonymous voter registration data, which also contained date of birth, zip code, and sex.  Because these three variables were available in both the de-identified data and the identified auxiliary data, Sweeney was able to merge them together (Figure \ref{fig:sweeney}).  Fortunately, rather than posting all of the records online, Sweeney mailed Massachusetts Governor William Weld a copy of his own records.  More importantly, Sweeney published a paper describing her attack and a proposing a defense against that kind of attack \citep{sweeney2002, sweeney2005, ohm2010}.

Beyond simple merges between two files, re-identification attacks can also involve multiple sources of auxiliary information.  For example, \citet{malin2004} combined databases of DNA records, which contained time and place of collection, with publicly available hospital discharge data.  This hospital discharge data in turn contained basic demographic information that could be used to combine it with identified voter registration records \citep{malin2004} (Figure \ref{fig:sweeney}).  In other words, the hospital discharge data served as a critical middle step as Malin and Sweeney linked the de-identified DNA records with identified voting records.

The hypothetical example of Lex Luthor clarifies that all data are potentially identifiable when facing a powerful adversary.  Further, Sweeney's re-identification attacks---as well as attacks by other privacy and security researchers \citep{narayanan2016}---show that re-identification attacks are possible in the real world and that these attacks benefit from the presence of rich auxiliary datasets.  These facts alone suggest that social scientists should be concerned about the possibility of re-identification attacks.  A further cause for concern is that re-identification attacks are probably easier today than at any time in the past.  Just as scientists are excited about big data sources for research, adversaries can use these same big data sources for re-identification attacks.  Because all data are potentially identifiable and because re-identification attacks are easier now than at any time in the past, we were concerned about the possibility of a re-identification attack during the Fragile Families Challenge.

\subsection{Potential technical solutions}
\label{technical}

Because the risk of re-identification attacks occurs in many situations, researchers have developed techniques that facilitate the analysis of data while protecting the privacy of the people described in the data \citep{willenborg2001,duncan2011,dwork2014}.  We believe that many social scientists, following the lead of statisticians, would organize these privacy-preserving techniques into two main groups: those that focus on modifications to the data and those that focus on access to the data \citep{reiter2011,duncan2011}.  However, our own deliberations were more influenced by the literature in computer science, and so we organized these privacy-preserving techniques into two different groups: those that offer provable guarantees and those that do not.  Techniques that offer provable guarantees would enable us to provide specific, mathematical bounds on what an adversary might be able to learn about individuals in the dataset, while making minimal assumptions about the adversary's knowledge, capability, or behavior.  During our deliberations, we considered two approaches that offer provable guarantees: differential privacy and cryptography.  Unfortunately, as we will now describe, we did not believe either approach was feasible in our setting.  For the Challenge we relied on many techniques that didn't offer provable guarantees.  Before describing the techniques we used, we will summarize the approaches we considered that offer provable guarantees, and we will describe why we did not think they were appropriate in our setting.

The first main approach that we considered that offers provable guarantees is differential privacy \citep{dwork2008,dwork2014}.  Differential privacy is a set of techniques for developing data release algorithms that achieve the following privacy guarantee: any output of the data release algorithm would have been roughly as likely even if any particular record had been removed from the data. Intuitively, this means that the adversary can't tell from the data release whether or not any individual's record was included in the sample, and this property is definitionally treated as a kind of individual privacy.    Within the broad area of differential privacy, we considered two families of approaches.  Under one family of approaches---called \emph{non-interactive approaches}---we would release a modified form of the data to Challenge participants.\footnote{The idea of releasing a modified form of the data in order to increase privacy protections is common in social science.  However, many of the approaches that are typically used with social data, such as top-coding and coarsening, do not generally offer provable guarantees within the framework of differential privacy.  This does not mean that these approaches should not be used. In practice, we think they often make re-identification attacks more difficult, and we used some of them as described in Section~\ref{modifications}.}  Under the other family of approaches---called \emph{interactive approaches}---we would not release any data; rather, we would keep the data on a secure server and allow researchers to send queries to the server.\footnote{Some researchers refer to these two families as \emph{offline approaches} and \emph{online approaches} \citep{dwork2014}.}  

Within the family of non-interactive approaches, we considered two sub-families of approaches: transformed data\footnote{Under the transformed data approach, we would not release individual-level data but rather some aggregated form of data that could then be used for analysis.  For example, for a movie recommendation task, \citet{mcsherry2009} argue that many predictive algorithms operate on the movie-movie covariance matrix, and they demonstrate how to achieve differential privacy for this class of algorithms by releasing a perturbed version of the covariance matrix. We did not believe that such an approach was possible in the Challenge because we are not aware of an aggregated data structure that would not substantially limit the modeling techniques that would be available to Challenge participants.} and synthetic data\footnote{Under the synthetic data approach, we would have to build a generative model which, when sampled, produces data from the same joint distribution as the Fragile Families Study data.  Further, we would have to build this generative model in a way that is differentially private.  There are two main approaches to building such models (differentially private or not): using domain expertise \citep{drechsler2011} and algorithmic learning \citep{hardt2012}.  We did not believe that we had sufficient social science domain expertise to create a realistic data generating process for the joint distribution of all 12,942 variables in the Challenge dataset.  To the best of our knowledge, existing applications of synthetic data created based on domain expertise generally involve a much smaller number of variables.  For example, the U.S. Census Bureau released a longitudinal dataset of businesses that contained 5 synthetic variables~\citep{kinney2011}.  Although algorithmic learning approaches do not require domain expertise, they too are generally limited to datasets with a small number of variables.  For example, a recent technique based on generative adversarial neural networks was applied to a clinical trial with 36 variables and 6,502 observations \citep{beaulieu2017}.  To the best of our understanding, this  and other related algorithmic approaches will not be effective in generating synthetic data with 12,942 variables.}, but we concluded that neither  approach was feasible in our setting.

In addition, we also considered interactive approaches where we would have hosted the Challenge data on a server and allowed Challenge participants to query the data (e.g., request a specific regression model).  We would then return results with carefully generated noise that would satisfy differential privacy.  We found that interactive approaches to differential privacy required major changes to the workflow of analysts and restricted the types of analysis that were possible.\footnote{An early tool for interactive differentially private data analysis is PINQ (Privacy Integrated Queries) \citep{mcsherry2009PINQ}. PINQ is geared toward data analysis and summary statistics like SQL (Structured Query Language). It also supports basic machine learning algorithms, but it is not clear if it allows building complex machine learning models with many predictors.  PINQ requires data analysts to learn a new programing language---a combination of C\# and LINQ, a SQL-like language---in order to express their queries.  Further, like all differentially private systems, PINQ imposes a \emph{privacy budget} whereby each query has a query-specific privacy cost and analysts have a fixed budget.  This budget ensures that the claimed differential privacy guarantees can be met, but it introduces substantial complexity for analysts who are not already familiar with differential privacy.  More recently, the Harvard University Privacy Tools Project is attempting to build techniques for differentially private access to social science datasets.  They have developed a tool named PSI (Private data Sharing Interface), but it did not appear to be publicly available at the time of the Challenge.  Further, based on a recent paper \citep{gaboardi2016}, it appears that PSI does not (yet) support high-dimensional statistical and machine learning models; it is not clear if these limitations are fundamental.  We believe that tools like PINQ and PSI are promising approaches to interactive differential privacy, and organizers of future projects similar to the Fragile Families Challenge should consider these tools and related tools that may be developed in the future.}  In conclusion, although we think differential privacy is an elegant and promising approach to providing provable guarantees, we did not think these approaches---either non-interactive or interactive---were feasible for the Fragile Families Challenge.\footnote{There are also other parts of the differential privacy literature that appear related to our problem but turn out not to be relevant.  For example, there is a large literature on techniques for differentially private machine learning. In this body of work, the privacy problem to be solved is that the model trained on private data (for example, a set of weights for logistic regression) might itself leak information about the training data. This research assumes that the learning algorithm has direct access to the raw data; the privacy question pertains to the algorithm's {\em outputs}. Thus, it is not applicable to our setting, as it assumes that researchers have access to the raw data.}  

The second major class of provable privacy techniques that we considered is based on cryptography.  Specifically, we considered ideas related to homomorphic encryption, which is a technique to encrypt data in such a way that computing a function $f()$ on the encrypted data and decrypting the result yields the same output as computing $f()$ on the original data.\footnote{More precisely, the code for $f()$ is transformed into a function that operates on encrypted data, and these operations are potentially computationally expensive, requiring cryptographic operations for every bit manipulation in $f()$.  Early homomorphic encryption techniques were limited in the variety of functions that could be computed under encryption, and so they were called ``SomeWhat Homomorphic Encryption'' (SWHE). A breakthrough by \citet{gentry2009} removed these limitations enabling ``Fully Homomorphic Encryption'' (FHE).  The downside is that FHE introduces computational overheads that make it currently impractical in many contexts.}  For the Challenge, we imagined utilizing homomorphic encryption as follows: (1) we could homomorphically encrypt the Challenge data and release it to everyone because the encryption would make re-identification attacks infeasible; (2) Challenge participants could build specially constructed models on the encrypted data (models designed to work on the unencrypted data cannot be run on the encrypted data); (3) participants could upload their encrypted predictions to the Challenge server; (4) the Challenge server would decrypt the predictions and calculate the mean square error.  While this approach sounds very promising, there a number of concerns---both conceptual\footnote{Conceptually, there is a much simpler way to offer the same level of privacy guarantees: releasing no raw data, requiring contestants to upload code to the Challenge server, executing that code on the server, and revealing only the mean squared error.  We think homomorphic encryption is useful only when the data owner is computationally limited or both parties have private inputs. In the former case, homomorphic encryption allows outsourcing of expensive computations (such as machine learning), which might be desirable even after we account for the slowdown introduced by computation under encryption. In the latter case, homomorphic encryption allows the data recipient to keep a proprietary algorithm secret.  Neither of these conditions were met in our situation.} and practical\footnote{Practically, at the time of the Challenge it took substantial expertise to create even extremely simple statistical models that could run on encrypted data.  Experts on homomorphic encryption have likened the process of creating these models to programming in assembly language \citep{crawford2018}.}---which led us to conclude that approaches using homomorphic encryption were not appropriate for the Challenge.

To summarize, we did not think that approaches that offered provable guarantees--- differential privacy or cryptography---were applicable \emph{in our setting}.  We hope that these techniques will continue to improve and will ultimately become more useful in this type of setting in the future.  Because we could not deploy a technique with provable guarantees, we undertook a process of threat modeling and threat mitigation, which we now describe.

\section{Threats and mitigations}
\label{threat_modeling_and_threat_mitigation}

Given that re-identification attacks were possible and technical solutions with provable guarantees were not available, we sought to better understand the possible risks and then reduce them as much as possible.  Therefore, we undertook a process of threat modeling and threat mitigation \citep{shostack2014}.  During the threat modeling we tried to imagine very specific, concrete risks.  When considering these risks, we found that it was helpful to separately consider the probability of harm and the magnitude of harm.  Further, we tried to avoid spending an inappropriate amount of time considering high harm events with a low probability \citep{sunstein2002}.  We also found it helpful to separate risk to us as organizers of the Challenge from risk to the survey respondents.  We were much more accepting of risk to ourselves (e.g., reputational risk) than risk for study participants.  Once these risks were identified, we tried to design steps that would mitigate these risks.

Our threat modeling was primarily focused on re-identification attacks and revolved around two main questions: (a) who might have the skills and rich auxiliary information that would be needed for a re-identification attack? and (b) who might have the incentive to carry out such an attack? 
In order to help answer these questions, we conducted an in-house attack of our own data.  We imagined possible data sources that could (a) be identified and (b) contain variables that also exist in the Fragile Families Study.  Just as \citet{sweeney2002} merged identified voting records with de-identified medical records in order to create identified medical records (\ref{fig:sweeney}), we tried to find information an attacker might merge with the Fragile Families Challenge data.

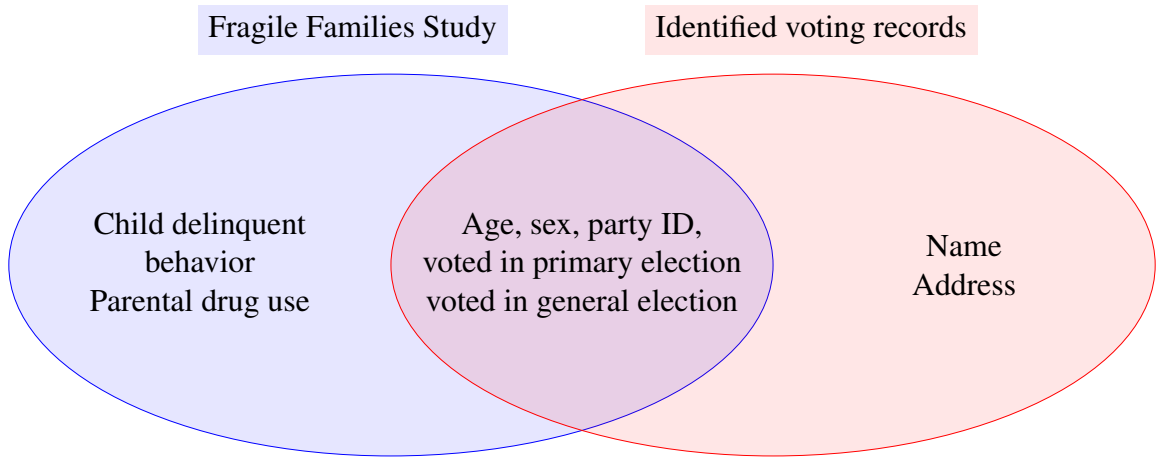
\begin{figure}
\centering
\begin{tikzpicture}[x = 2in, y = 1in]
\draw[blue, fill=blue, fill opacity = .1, line width = 0pt] (0.5,0) ellipse (2in and 1in);
\draw[red, fill=red, fill opacity = .1, line width = 0pt] (1.5,0) ellipse (2in and 1in);
\node[align=center] at (0,0) {Child delinquent\\behavior\\ Parental drug use};
\node[align=center] at (2,0) {Name\\Address};
\node[align=center] at (1,0) {Age, sex, party ID,\\ voted in primary election\\ voted in general election};
\node[align=center,fill = blue, fill opacity = .1, text opacity = 1, anchor=south] at (0.4,1.1) {Fragile Families Study};
\node[align=center,fill = red, fill opacity = .1, text opacity = 1, anchor=south] at (1.6,1.1) {Identified voting records};
\end{tikzpicture}
\caption{\textbf{Hypothetical example of re-identification attack of the Fragile Families Study.} The Fragile Families Study does not contain information on voting, but if it did, these variables could be linked to an identified auxiliary data source: administrative voting records.  After completing this linkage, an adversary could learn about potentially sensitive information, such as parental drug use and child delinquent behavior.  This hypothetical attack, as well as our actual in-house attack, would be possible even if the data did not include any geographic information.}
\label{fig:vennDiagram}
\end{figure}

After envisioning and investigating many types of auxiliary data, we used one such source to attack our own data. This in-house attack demonstrated that many respondents were unique in both the Fragile Families Study and the auxiliary data source, and it led to modifications of our data that we will describe later in the paper.  We will illustrate the general structure of our attack with a hypothetical example.  Suppose that the Fragile Families Study collected data on voting behavior, recording in each wave of data collection variables such as sex, age, party ID, and whether the respondent voted in the most recent primary and general elections.  These variables are also available in identified voter registration databases which would allow an adversary to merge them (see Figure \ref{fig:vennDiagram}).\footnote{The ability of this hypothetical merge to succeed depends on how unique people are in the dataset and in the population.  As an extreme example, imagine a child's mother who was 24 years old in 2000. Suppose this mother registered with the Green party in 2000, the Republican party in 2002, and the Democratic party in 2004. In each year, she voted in the primary election but not the general election. This respondent's particular combination of variables may be unique in the entire U.S. population and this person would be at risk in this kind of merge.  More generally the more unique each person is, the more vulnerable they would be to this kind of attack.}  A key feature of this hypothetical attack, as well as our real in-house attack, was that it did not involve the use of zip code or any other geographic information.  Thus, even though the two real attacks that we described previously in this paper used zip code (Figure \ref{fig:sweeney}), we wish to highlight the fact that it is still possible to re-identify data even if it does not contain obvious geographic information.

The process of attacking our data was useful for three reasons. First, it helped us realize how a small number of variables could substantially aid re-identification.  In particular, continuous variables like age made re-identification especially easy because they differentiated people into many groups. We decided to redact or modify variables that we thought were most likely to be the target of an attack, as we will describe more below.  Second, our attempts to attack our data made us realize the difficulty of obtaining auxiliary data at the national level. An attacker to the Fragile Families Challenge file would likely need a dataset that is national in scope, but state differences in data sources make assembling such a data source difficult (but certainly not impossible).  Finally, attempting to attack our own data made our fears about re-identification concrete and produced a clear way to explain the risks to our Board of Advisors and to other stakeholders.

This in-house re-identification attack was part of a larger process of moving away from a general fear of re-identification toward specific, actionable worries about particular people with a dangerous combination of capability and incentives.  Next, we summarize the five biggest threats on which we focused, roughly ordered by the amount of risk we think they posed. Then, we briefly discuss other threats on which we did not focus during the Challenge but which might be important in other settings.  After describing these threats, we describe the six main steps we took to mitigate those threats, roughly ordered by our perception of their importance (Table \ref{tbl:threats}).  When describing our threat modeling and threat mitigation process, we will be intentionally vague at certain points because versions of the Fragile Families Study data exist in the research community outside of our control, and we do not wish to increase the risk of a re-identification attack in the future.

\begin{table}
\resizebox{\textwidth}{!}{
\begin{tabular}{lVVVVVV}
& Low & Careful & Challenge & Application  & Ethical & Modifications\\
& profile & language & structure & process & appeal & to data \\
\hline
Privacy researcher & \checkmark & \circlecheckmark & \checkmark & \checkmark & \checkmark &  \\
Nosy neighbor & \circlecheckmark & & & \checkmark & & \\
Troll & \circlecheckmark &  & \checkmark & \checkmark &  & \checkmark \\
Journalist & \checkmark & \checkmark & \checkmark & \circlecheckmark & \checkmark & \checkmark \\
Cheater &  & \checkmark & \circlecheckmark &  & \checkmark & \checkmark
\end{tabular}
}
\caption{\textbf{Five main threats and the six main steps we took to mitigate those threats.} Rows represent potential threats. Columns represent actions we took to mitigate the threats. Check marks indicate that we expected the action to be effective against the adversary. For each adversary (row) the circled check mark represents the action we felt was most effective against that adversary.}
\label{tbl:threats}
\end{table}

\subsection{Threats}

\subsubsection{Threat 1: A privacy researcher}
\label{privacyresearcher}

Privacy researchers represent an important ``threat'' to any social science dataset.  They have the skills to re-identify the data and the incentives to conduct and publicize an attack.  Although we describe privacy researchers as a ``threat,'' we wish to emphasize that these researchers have good intentions.  Some privacy researchers undertake attacks with the goal of developing defensive techniques that can prevent future attacks.  Other privacy researchers might seek to illustrate privacy problems in a particular dataset (e.g., \citealt{zimmer2010}), with the goal of encouraging other data stewards to be more careful before a true adversary finds the problems.\footnote{Based on our experience, we believe that this strategy is effective.  The threat of privacy researchers caused us to be more careful.}  

Because privacy researchers represent an important and sometimes misunderstood threat, it is helpful to briefly describe the history and norms of this community so that other researchers can better understand their motivations.  Today's information security and privacy research community traces its intellectual lineage in part to the field of cryptography and its military applications in which secrets must be defended against powerful adversaries, with human lives at stake \citep{kahn1996}. The research culture of cryptography is shaped by its painful history of naively optimistic claims of ``unbreakable'' ciphers. Centuries of failures of such claims gradually established the importance of adversarial analysis as a scientific technique. Computer scientists today believe that scientific rigor in data privacy protection technology can be obtained only if claims are subject to adversarial scrutiny \citep{menezes1996}.  Further, in the absence of a specific known adversary, privacy researchers believe that the prudent course of action is to assume the {\em union} of all such adversaries, i.e., a very powerful one.  Assuming a capable adversary with complete knowledge of the system is a central principle in cryptography known as Shannon's maxim \citep{shannon1949}.

Privacy researchers consider real data releases to be valuable targets for demonstrations (i.e., \citealt{sweeney2002,narayanan2008,zimmer2010}), because they believe that work with toy data sets has limited ecological validity.  Apart from the scientific benefit, re-identification demonstrations on real datasets are seen as ways to warn consumers of risks and disincentivize bogus claims of security.  However, not all real data releases are equally attractive targets, at least for academic privacy researchers.  It is difficult to publish re-identification research unless there is novelty in the method, and this desire for novelty has served as a check on the number of such studies in practice. In other words, ironically, datasets that are too easy to re-identify are likely to escape the attention of privacy researchers. 

What about the risks of re-identification research?  Debate around this question is informed by the debates on the ethics of offensive computer security research more generally.  In short, this community has concluded that they must engage in privacy attacks in order to anticipate and mitigate weaknesses in data protection before they are discovered by more nefarious adversaries.  





\subsubsection{Threat 2: A nosy neighbor}

In addition to privacy researchers, a very different kind of threat comes from a group of people with very different motivations and knowledge, a group often called \emph{nosy neighbors}.  These people often have a specific interest in a single person in the dataset and already have substantial auxiliary information about that person.  In the Fragile Families Challenge, we imagined that a mother might want to re-identify the data to learn about the survey answers provided by the father.\footnote{The threat of a nosy neighbor attack can be illustrated through the example of the Netflix Prize, a mass collaboration that partially inspired the Fragile Families Challenge and which was subject to a re-identification attack \citep{narayanan2008}.  In 2006, Netflix offered a million-dollar prize to the team that could most improve its movie recommendation algorithm. Netflix released a dataset of ratings made on specific movies by specific users.  Researchers were challenged to use the ratings in the public data to predict held-out movie ratings. Before Netflix released the data, they took some steps to prevent re-identification and the data appeared to be anonymized; they consisted solely of movie ratings without any explicit individual identifiers. However, \citet{narayanan2008} found that the vast majority of users had histories of movie ratings that were unique in the sample and were statistically likely to be unique in the population. If you knew some of the movies one of these individuals had watched, you could re-identify their row in the data and see \emph{all} of the movies they had rated. As the authors describe, ``a water-cooler conversation with an office colleague about her cinematographic likes and dislikes may yield enough information'' \citep{narayanan2008}. Given this water cooler information, a nosy neighbor would only need minimal technical abilities to re-identify the colleague by retrieving the user record that most closely matched the known likes and dislikes.}  The linked nature of the Fragile Families data makes a nosy neighbor attack easier because the attacker could herself be in the data.  They would only need to find themselves in order to learn more about the responses of the people in their family.

\subsubsection{Threat 3: A troll}
\label{troll}

Third, we considered the possibility of a \emph{troll} who might attempt to re-identify the data because they enjoy causing trouble or seeking attention (see \citealt{phillips2015}).  While some adversaries (i.e., a privacy researcher) might be able to harm our academic careers, a troll who posted respondents' information online might actually harm survey respondents. 

We also considered the possibility of a ``hacktivist'' who might attempt to re-identify the data to make a larger political point. For instance, the hacker group Anonymous publicly posted the names and social media profiles of members of the KKK in 2015. When doing so, they wrote: ``We hope Operation KKK will, in part, spark a bit of constructive dialogue about race, racism, racial terror and freedom of expression'' \citep{FranceschiBicchierai2015}.  Might there exist adversaries who have similarly negative feelings toward social scientists doing research on a disadvantaged population? We would hope that potential hacktivists would recognize our good intentions, but we could not rule out the possibility that someone would attack the study to make a statement against our project or against social science research in general.

\subsubsection{Threat 4: A journalist}

A fourth adversary we considered was a journalist. For example, in 2006 AOL made the searches of thousands of users available to the public, assuming that one's search terms would not be easily traceable back to an individual's identity. Contrary to their expectations, \citet{barbaro2006} wrote a widely-read \emph{New York Times} article revealing the identity of one woman whose search information had been included in the release. We worried that a journalist could do the same thing with the data from the Fragile Families Challenge. Like a privacy researcher, a journalist might be motivated to attack the data to make a larger point. However, we believe that a journalist bound by the norms of his or her profession would be unlikely to intentionally harm study participants.  Therefore, we reasoned that journalists posed a greater risk to us, as organizers, than to the survey respondents.

\subsubsection{Threat 5: A cheater}
\label{cheater}

Finally, prior challenges have been won through strategies involving re-identification \citep{narayanan2011}. We worried that, if we set up the Challenge with a big prize and no clear prohibitions on linkage to auxiliary data, someone might try to win by re-identification. An adversary who re-identified the respondents could contact them and discover their outcomes, thereby achieving remarkably successful predictive performance.

\subsubsection{Other threats}

These five adversaries---privacy researcher, nosy neighbor, troll, journalist, and cheater---were the ones that we considered most closely in our threat modeling.  However, this list is not exhaustive of the threats that we considered or the threats that might arise in other situations.  For instance, three other adversaries that we considered---and which might be more important in other settings---are governments, criminals, and companies. 

Certain parts of the U.S. government most likely have the skills and rich auxiliary data that would be needed for a re-identification attack of our data.  They might also have the incentive if our data contained information they could not find elsewhere, such as if respondents in our survey had reported on their experience as spies for a foreign government.  Because we deemed the information in our data to be of little value to the U.S. government, we did not believe the U.S. government had an incentive to conduct such an attack in our setting.

A different set of attackers might be motivated by financial gain.  For example, companies seeking to build databases for targeted marketing might try to acquire large social science datasets.  Given the size, structure, and de-identification of our data, however, we believed that it would be unattractive to companies.  Further, sophisticated criminals might wish to attack a dataset containing credit card numbers or containing compromising information on wealthy individuals that could be used as the basis for blackmail.  Our data do not contain information like credit card numbers, and we reasoned that an adversary with the goal of finding compromising information on elites would more likely target other datasets.  Therefore, we believed that an attack motivated by financial gain---either by a company or criminal---was unlikely in our setting.

While our threat modeling was mainly focused on re-identification attacks that would occur through a merge with auxiliary information, other attacks were also possible.  For example, we considered that someone might attempt to learn the identity of the survey respondents by breaking into the Fragile Families Study offices and physically stealing computers.  We deemed this possibility extremely unlikely, mostly because we did not see a clear incentive to carry out this attack.  Further, there are defenses in place that would make this attack more difficult than it appears.  These defenses also make the possibility of accidental leak of information extremely unlikely.  Overall, we would recommend that other researchers conduct a similar threat modeling exercise, keeping in mind that the threats in each situation might be different.

\subsection{Threat mitigation}
\label{mitigation}

There is no way with present technology to completely eliminate the risks that these threats pose while achieving the scientific objectives of the mass collaboration.  Nevertheless, we took six main steps to make an attack more difficult and less attractive. The columns of Table \ref{tbl:threats} represent these steps and their expected efficacy against various adversaries. We have ordered our actions in terms of our perception of their importance, from most to least important.

\subsubsection{Low profile}
\label{low}

When organizing a mass collaboration, one might seek press coverage in major national and international venues.  Such publicity would help to attract the widest and most diverse pool of participants possible, but it could also increase the risk of attack. High-profile studies are more likely to attract the attention of a nosy neighbor, a troll seeking attention, or a journalist or privacy researcher looking to re-identify a project that will draw interest from a wide readership. A low-profile study is less likely to be noticed by these adversaries and might be a less attractive target. 

During the Challenge, we decided to keep a relatively low profile, and we focused our outreach on settings with a high probability of yielding participants who could contribute and a low probability of yielding participants who might attack the data. 
 
For example, to raise awareness about the Challenge, we emailed the directors of Population Centers funded by the National Institutes of Health, and we hosted getting-started workshops at universities, in courses, and at scientific conferences. Our strategy of keeping a low profile can be easily adopted by other mass collaborations.

\subsubsection{Careful language}
\label{careful}

Many data stewards may not realize it, but using careful, precise, and humble language may help prevent an attack from a privacy researcher or journalist. For instance, \citet{zimmer2010} made an ethical example out of the Tastes, Ties, and Time study in part because the original authors made strong statements such as ``all identifying information was deleted or encoded,'' (quoted by \citealt{zimmer2010} from the original study codebook). Privacy researchers may wish to correct data stewards who make overly confident statements about the de-identified nature of their data. By choosing language carefully, data stewards can be more honest about the safety of their data, thereby removing the need for privacy scholars to correct them. Instead of saying that ``all identifying information was deleted,'' one might say, ``we removed information that was obviously identifiable.'' Instead of writing that data are ``anonymized,'' data stewards should write that steps were taken to make the data less identifiable. Small changes in language can help to convey that data stewards understand the privacy risks and have made a reasoned judgment to proceed anyway. For example, we wrote the following on the Fragile Families Challenge website: 
\begin{quote}
``All participants in the Fragile Families and Child Wellbeing Study have consented to have their data used for social research. These procedures, as well as procedures to make de-identified data available to researchers, have been reviewed and approved by the Institutional Review Board of Princeton University (\#5767). The procedures for the Fragile Families Challenge have been reviewed and approved by the Institutional Review Board of Princeton University (\#8061). In addition, we have also taken further steps to protect the participants in the Fragile Families and Child Wellbeing Study. If you would like to know more, please send us an email.''
\end{quote}
We believe that the relatively easy step of using careful language and inviting contact from potential attackers decreased the risk of attack.

\subsubsection{Structure of Challenge}
\label{structure}

The structure of the Challenge was also designed in part to decrease the incentives to attack the data.  By avoiding asking Challenge participants to predict sensitive outcomes, such as involvement in the criminal justice system or sexual behavior, we think that we reduced the risk of attack from a privacy researcher, journalist, or troll.  Avoiding potentially sensitive outcomes was part of keeping a low profile, and we suspect that it was important. 

We also built certain things into the Challenge that would decrease the risk of a cheater attacking the data.  First, in contrast to other high-profile challenges that offered large monetary prizes \citep{bennett2007}, we chose to reward the best submissions with a trip to Princeton University to discuss their work. This prize was designed to emphasize an intrinsic goal of knowledge creation rather than an extrinsic financial incentive, thereby reducing incentives to cheat.  Second, we emphasized in all promotional materials that the Challenge was a mass collaboration and a new way of working together, not a competition. For example, the banner at the top of our website read, ``What would happen if hundreds of social scientists and data scientists worked together on a scientific challenge to improve the lives of disadvantaged children in the United States?'' In the approval process, we also asked people about their motivations to participate, thereby encouraging them to reflect on their reasons. Nearly all participants responded with motivations that involved participation in scientific research or helping disadvantaged children.  Third, we made the rules of the Challenge very clear that linking to auxiliary data sources was not allowed, and that anyone who did so would be disqualified.  Fourth, we required all participants to upload their code and a narrative explanation along with their prediction.  If a cheater successfully attacked the data and merged in outside information, this process would have to be obscured in whatever code was uploaded.  Altogether, we think that these aspects of the structure of the Challenge decreased incentives to attack the data.

\subsubsection{Application process}
\label{application}

In the interest of open and reproducible science, many have argued that research data should be made public. While we agree with the spirit of this call, we join others making the more modest call for open sharing of data with allowable constraints when privacy or other concerns must be balanced against the goal of open science \citep{freese2017,salganik2018}. In particular, an application process can help ensure that only those who can plausibly yield scientific benefit be given access to the data and that participants who pose increased risk can be monitored more closely.

\begin{figure}
\centering
\begin{tikzpicture}[x = 3in, y = .8in,every node/.style={font=\footnotesize}]
\node[align=center,draw,rounded corners,fill=blue,fill opacity=.05,text opacity = 1] (apply) at (0,.5) {Application received};
\node[align=center,rectangle,draw,yscale=7,xscale = 12,rotate=45,fill=yellow,fill opacity=.1,text opacity = 1] (review1) at (0,-1) {};
\node[align=center] at (0,-1) {Central organizer\\(Lundberg or Salganik)\\perceives a threat?};
\node[align=center,anchor=south] at (.5,-1) {Yes};
\node[align=center,anchor = east] at (0,-2) {No};
\node[align=center,draw,rounded corners,fill=blue,fill opacity=.05,text opacity = 1] (flags) at (1,-1) {Organizer follows up\\with applicant};
\node[align=center,rectangle,draw,yscale=3,xscale = 10,rotate=45,fill=yellow,fill opacity=.1,text opacity = 1] (response) at (1,-2) {};
\node[align=center] at (1,-2) {Applicant responds?};
\node[anchor = west] at (1,-2.5) {Yes};
\node[anchor = south] at (1.35,-2) {No};
\node[align=center,draw,rounded corners,fill=blue,fill opacity=.05,text opacity = 1] (noResponse) at (1.6,-2) {End of process};
\node[align=center,rectangle,draw,yscale=3,xscale = 10,rotate=45,fill=yellow,fill opacity=.1,text opacity = 1] (realThreat) at (1,-3) {};
\node[align=center] at (1,-3) {Deemed a threat?};
\node[align=center,anchor=south] at (.5,-3) {No};
\node[align=center,anchor=west] at (1,-3.5) {Yes};
\node[align=center,draw,rounded corners,fill=blue,fill opacity=.05,text opacity = 1] (local) at (1,-4) {Applicant invited to participate\\locally on a secure computer};
\node[align=center,draw,rounded corners,fill=blue,fill opacity=.05,text opacity = 1] (recommendation) at (0,-3) {Recommendation for approval sent\\to full review committee};
\node[align=center,rectangle,draw,yscale=4,xscale = 12,rotate=45,fill=yellow,fill opacity=.1,text opacity = 1] (approved) at (0,-4) {};
\node[align=center] at (0,-4) {Approved by committee?};
\node[align=center,anchor = east] at (0,-4.5) {Yes};
\node[align=center,anchor = south] at (0.5,-4) {No};
\node[align=center,draw,rounded corners,fill=blue,fill opacity=.05,text opacity = 1] (terms) at (0,-5) {Participant completes Terms and Conditions\\and agrees to behave ethically};
\node[align=center,draw,rounded corners,fill=blue,fill opacity=.05,text opacity = 1] (data) at (0,-6) {Data sent to participant\\through secure online file sharing};
\node[align=center,draw,rounded corners,fill=blue,fill opacity=.05,text opacity = 1] (call) at (0,-7) {Participant calls to request password to open the data};
\draw[->, line width=2pt] (apply) -- (review1);
\draw[->, line width=2pt] (review1) -- (flags);
\draw[->, line width=2pt] (flags) -- (response);
\draw[->, line width=2pt] (response) -- (noResponse);
\draw[->, line width=2pt] (response) -- (realThreat);
\draw[->, line width=2pt] (realThreat) -- (local);
\draw[->, line width=2pt] (realThreat) -- (recommendation);
\draw[->, line width=2pt] (review1) -- (recommendation);
\draw[->, line width=2pt] (recommendation) -- (approved);
\draw[->, line width=2pt] (approved) -- (terms);
\draw[->, line width=2pt] (approved) -- (local);
\draw[->, line width=2pt] (terms) -- (data);
\draw[->, line width=2pt] (data) -- (call);
\end{tikzpicture}
\caption{\textbf{Screening process for applications to the Fragile Families Challenge.} Each potential participant completed an application describing their educational background, research experience, and motivations to participate in the study. We assessed these applications in terms of their ability to contribute to the goals of the Challenge and in terms of the risk that an applicant might try to re-identify respondents.  Each application was reviewed by one of the lead organizers of the Challenge, sent to a review committee, and then approved 24 hours later if there were no objections. After approval, participants completed a set of terms and conditions, received a link to an encrypted file, and then called us for a password to open the file. For further details, see the main text. To review the application form and the terms and conditions, see the Appendix.}
\label{fig:screening}
\end{figure}
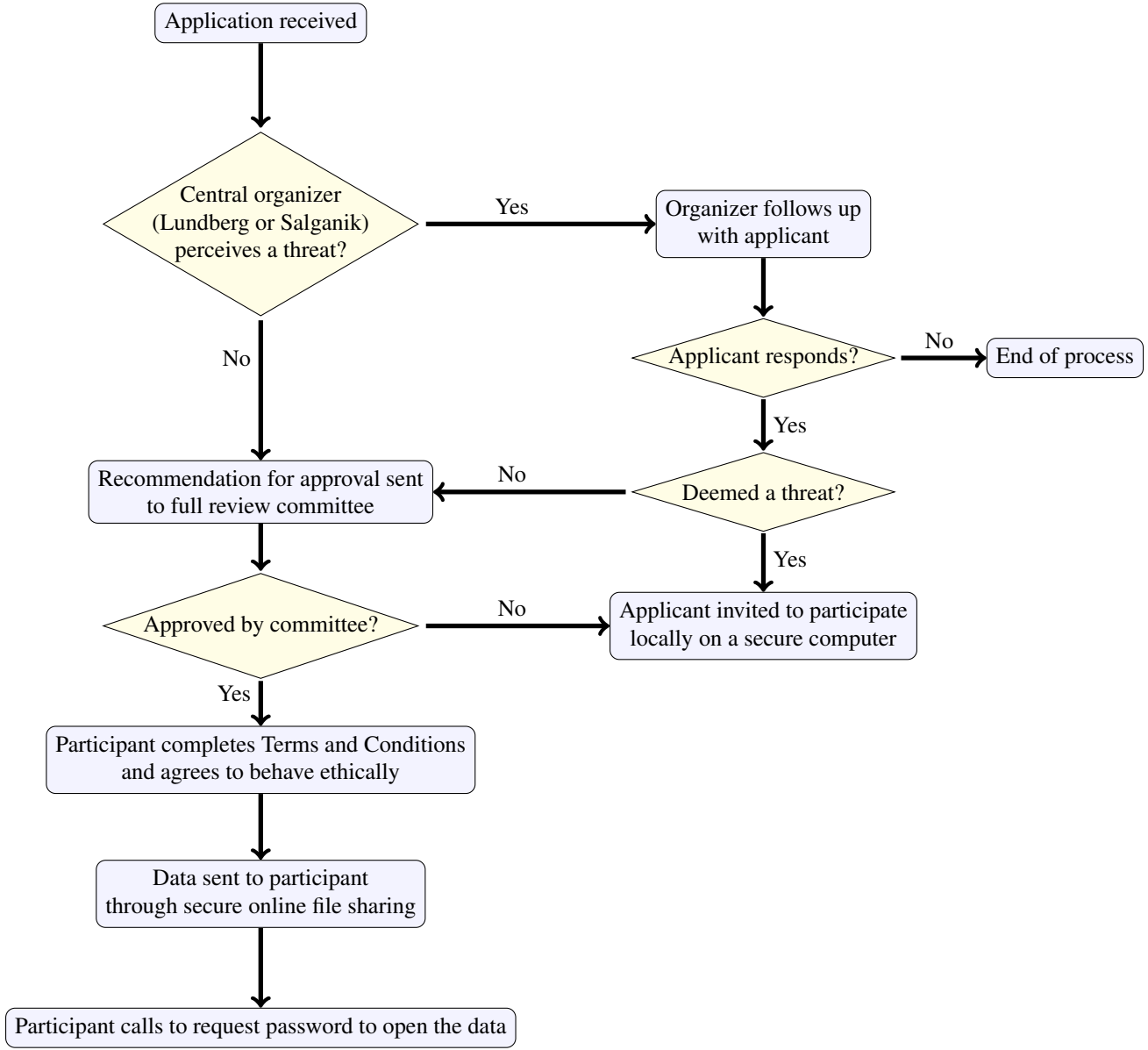

For the Challenge, we developed a process to screen applicants (Figure \ref{fig:screening}).  Initially, people interested in participating completed an application that asked for information about their educational background, research experience, and motivations to participate in the study (see Appendix for the exact application). Responses to the application were sent to an email account checked by two of the Challenge's central organizers (Lundberg and Salganik). One of the Challenge organizers would provide an initial review of the application, in many cases searching the internet to corroborate claims made in the application or to look for important information that was excluded.\footnote{Applications from students were sometimes difficult to evaluate, given that students often have little research experience to report. We occasionally spoke with a student's academic advisor to ensure that participation was being overseen by a responsible individual who understood the importance of respecting the data. In one case, we pointed a student toward alternative datasets that were less sensitive which would serve as equally useful for that student's project.} The reviewer would make an initial recommendation as to whether a participant should be approved and would send a summary of the application to a team of eight reviewers including all the authors of this paper, survey administrators, and the Principal Investigators of the Fragile Families Study. Members of this broader review team were given 24 hours to raise any opposition to the application.\footnote{We occasionally omitted this waiting period when individuals were participating in a known setting, such as a class assignment or a workshop in which we spoke directly with potential participants.}  

After this review process, we required approved applicants to type a set of statements acknowledging their agreement with our terms and conditions (the full set of statements is provided in the Appendix).  The purpose of the terms and conditions was not to screen participants, but to make sure they understood their ethical responsibilities (see Section \ref{appeal}).

After an approved applicant had agreed to the terms and conditions, we emailed her a link to an encrypted data file (the link automatically expired within seven days).  Finally, the approved participant had to call us by phone to receive the password to decrypt the file. This last step reduced the risk that an email could be intercepted and also provided an opportunity to speak in person with participants and thereby reinforce that the project was a mass collaboration involving real people, not a competition to be won at all costs.

During the Challenge, we received 472 applications\footnote{The first 103 of these applications came in the pilot test we ran in an undergraduate machine learning class at Princeton and followed a slightly different format from the rest because we modified the application procedure based on feedback from the pilot test.}, and most were uneventful. However, about 10 applications raised yellow flags that we addressed either by sending a follow-up email or by having a discussion with the applicant (in person or by video chat).  These conversations helped ensure mutual understanding of the importance of respecting the privacy of respondents. For instance, one applicant told us they were working on the ``record linkage problem with massive data.'' Another applicant told us of a research interest in re-identification. In both cases, we were glad to have spoken with the applicant before providing the data.  In one case, a researcher who had previous experience doing re-identification attacks did not respond to our follow-up email and was therefore not granted access to the data.

One might worry that this screening process---which involved many hours of work from the Challenge organizers, graduate students, study staff, and Principal Investigators---was completely futile because people could just lie in their applications.  However, from our threat modeling we suspected that well-intentioned privacy researchers might represent our greatest threat. These individuals believe strongly in ethical use of data, and we suspected that they would be honest in their applications. Because those who posed the greatest threat also seemed unlikely to lie in an application, we believe our screening process was a worthwhile endeavor.

One final aspect of the application process is worth describing.  Initially, we were worried about what might happen if we actually rejected an applicant.  Might this rejection anger someone and turn them into an even more motivated adversary who would attempt find the data elsewhere (i.e., from another participant) and then attack it?  Because of this concern, we made preparations to accept high-risk applicants as \emph{local participants}.  For these local participants, we prepared a special, secure computer in the office of the Fragile Families project director.  This computer was configured such that it was difficult to bring data onto the computer or take data off the computer (e.g., drives were deactivated and the computer was not connected to the internet and was locked to the wall).  However, we maintained the ability to re-connect this computer to the internet ourselves to upload a Challenge participant's predictions if needed.   Fortunately, we never had to use this secure computer, but it provided an opportunity for us to remain open to the potential participation of even the riskiest applicants.

\subsubsection{Ethical appeal}
\label{appeal}

It is not possible to force people to act ethically, nor is it possible in a screening process to determine with full confidence whether an applicant plans to act ethically. However, one can clarify for participants the ethical implications of misusing the data, thereby creating the possibility that those who might plan to re-identify the data will consider this plan from an ethical perspective before proceeding. After participants' applications were approved, they completed a set of terms and conditions designed to achieve this goal. Every participant read the following statement, written in bold, before participating.
\begin{quote}
``The Fragile Families and Child Wellbeing Study is a dataset of real people who have selflessly opened up their lives to us for the last 15 years so that their experiences can contribute to scientific research. By participating in the Fragile Families Challenge, you become a collaborator in this project. It is of the utmost importance that you respect the families in the data by using what they have told us responsibly.''
\end{quote}
After reading the statement, each participant typed a set of statements defining how they would use the data responsibly (see Appendix for specific statements). Participants could easily lie in this section, so it would not have stopped someone determined to act unethically. However, by outlining our expectations, we clarified a positive vision of what ethical behavior would entail.

\subsubsection{Modifications to data}
\label{modifications}

The final step we took to mitigate threats was to modify the data in the Challenge file.  Overall, these modifications were relatively minor because the preexisting Fragile Families Study files had already undergone an extensive process to promote privacy (Figure \ref{fig:versions}).  

First, as described earlier, we attempted to attack the data ourselves, and this led us to make certain changes in the structure of the background data file for the Challenge.  We choose not to describe these changes fully, but we note that some were inspired by prior re-identification demonstrations. We did not implement a $k$-anonymity approach \citep{sweeney2002}\footnote{$k$-anonymity states that data should only be released if each row is exactly the same as at least $k-1$ other rows, so that an adversary with perfect auxiliary data would be able to link a given individual with no fewer than $k$ records \citep{sweeney2002}. The higher the value of $k$ chosen, the more difficult it would be for an adversary to find all candidates and discover the true match. $k$-anonymity can be achieved by suppressing information on key features of certain individuals that would otherwise make them unique \citep{sweeney2002b}.  Unfortunately, $k$-anonymity quickly becomes infeasible as the number of features grows, since rows are rarely exactly the same on the joint set of all features \citep{aggarwal2005}, which means that a pure $k$-anonymity approach would substantially harm the utility of the data \citep{brickell2008}.  Despite the impossibility of applying $k$-anonymity to all 12,942 variables, we did apply the spirit of this approach to try to reduce the number of participants who were unique or nearly unique in the sample on a subset of variables.}, but we did make more minor changes such as adding noise to key variables to make a simple 1-to-1 merge more difficult.\footnote{The noise we discuss here is distinct in two ways from the noise added in differential privacy. First, we added noise to the inputs to participants' models, whereas differential privacy adds noise to the outputs.  Second, we added normally distributed noise, a choice distinct from the differential privacy approach of heavy-tailed Laplacian noise.  Our approach to adding noise does not yield provable privacy guarantees.}  Although adding noise has a long tradition in the statistical literature \citep{kim1986,reiter2012}, it is not foolproof in the high-dimensional setting. \footnote{For example, Netflix added noise to some of the data used in the Netflix Prize, but this did not prevent a re-identification attack \citep{narayanan2008}.}  Despite its limitations, adding noise to key variables at least makes a simple 1-1 merge with auxiliary data more difficult. For instance, we worried that height, weight, and body mass index (BMI) could be key linking variables if there was a breach of identified medical records data in the future, so we added a small amount of noise to these variables. We also added noise to other variables which we choose not to identify.  The noise we added came from an independent draw for each variable, with the exception of variables that measured the same construct (i.e., height in inches and height in centimeters). In these cases, we drew one noise term per child per construct, so that an adversary could not reduce the size of the noise by averaging over multiple responses.  Overall, we added noise to a few hundred variables.  To minimize the risk of unnecessary harm to the scientific promise of the project, the noise we added was always relatively small,\footnote{For more on the risks of adding too much noise see \citet{brickell2008}.} and Challenge participants were generally unaware of the added noise.\footnote{In one case a participant noticed that the distribution of some variables did not match that listed in the official study documentation, thereby correctly realizing that we had changed something in those variables. This participant emailed us, and we were happy to explain what we had done.} We suspect that adding noise made re-identification slightly harder with minimal harm to the scientific utility of the data.

In addition to seeking to make a re-identification attack more difficult, we also sought to reduce the magnitude of harm should re-identification occur by redacting some information that we thought might pose a serious risk to survey respondents.  When considering whether to redact information for harm reduction, we tried to weigh how much harm might come to how many people against how much the redaction might jeopardize the scientific goals of the project.  While making these decisions, we were aware that in a large dataset it can be difficult to know which information will be sensitive \citep[Ch. 6]{salganik2018}.

Finally, we decided that we did not want the Challenge data files to include anything that might obviously provide information about the survey respondent's location.  Unlike adding noise, we think this decision to strip geographic information may have decreased Challenge participants' ability to predict the six key outcomes.  However, we think that removing obvious geographic information made a re-identification attack much more difficult.  If an attacker were able to place a group of survey respondents in a particular city, then the attacker would only need an identified auxiliary data source that includes that city (e.g., state voting records).  However, if the respondents could be anywhere in the country, then the attacker would need national data which may be harder to acquire.  

Although many privacy audits focus mostly on modifications to the data, we saw this as just one of the six steps that we took to mitigate threats.  In fact, we think that many of other steps, such as keeping a low profile, were more important in this setting.

\subsection{Comprehensive assessment}
\label{assess}

After implementing the threat modeling and threat mitigation described above, which evolved over a period of months, we stepped back and comprehensively assessed who might have incentives to attack the data and what barriers might stand in their way (see Table \ref{tbl:threats}).

A privacy researcher would have to learn and care about the project despite our low profile. Even then, privacy researchers would only have an incentive to attack the data if they wanted to show publicly that we had done something wrong; our careful language incorporating the findings of privacy researchers mitigated against this danger. The privacy researcher would then have to lie in the application process or otherwise convince us they did not intend to re-identify the data, and they would have to proceed in the face of our ethical appeal. To the extent that privacy researchers are motivated by an ethical call for researchers to recognize the limits of privacy and handle data with care, we expected that these steps substantially reduced the risk that a privacy researcher would re-identify the data.

Nosy neighbors are almost impossible to stop with technical barriers; enormous changes to the data would be needed to render a respondent unrecognizable to a nosy neighbor. By keeping a low profile, we reduced the likelihood that a nosy neighbor would learn about the Challenge. By screening participants, we increased the chance that we would notice a nosy neighbor before sharing the data. With these changes in place, we determined that the risk of nosy neighbors increased only negligibly from the Challenge, compared with the risk that already existed from the availability of the Fragile Families Study data to researchers.

A troll might lie through the application process, be undeterred by an appeal to ethics, and have the technical skills to overcome our modifications to the data. However, a troll is likely to pursue the highest-profile targets for attention, so a low profile reduces the chance of a troll attacking the data.

A journalist would primarily be stopped by the application process: we expect that a journalist, because of the norms of their field, would not lie when applying to use the data.  Even a journalist who made it through the screening process would have to find the project interesting despite its low profile, ignore our appeal to ethics, and succeed in the technical difficulties of re-identification despite our modifications to the data. This seemed unlikely.

The threat from a cheater was primarily mitigated by the Challenge structure, which explicitly stated that anyone who cheated would be ineligible for a prize, which was non-monetary to begin with. With this structure, we doubted anyone would re-identify the data for the sole purpose of winning the Challenge.

To summarize, the design of the Challenge did not completely eliminate any of the threats we imagined, but it mitigated them to a sufficient degree that we concluded the Challenge only slightly increased the risk to participants above that which already existed from the use of the Fragile Families Study data by independent researchers.  

\section{Response plan}
\label{responsePlan}

Having mitigated---but not eliminated---the risk of re-identification, we created a response plan in case something went wrong.  More specifically, we took steps to ensure that we had the appropriate people involved in the project so that we could begin responding to a crisis quickly and forcefully.  Our Board of Advisors included a computer scientist who had previously re-identified several datasets (Arvind Narayanan), a sociologist and lawyer with expertise in data privacy, surveillance, and inequality (Karen Levy), and a journalist (Nicholas Lemann). In the event of an attack from a privacy researcher, the organizers of the Challenge would consult with Narayanan when responding. If a social scientist or policymaker criticized the project for mishandling the private information of disadvantaged children, the organizers of the Challenge would consult with Levy when responding. If the data were re-identified by a journalist who planned to publicize the story, the organizers of the Challenge would consult with Lemann. Each of these individuals is respected in various communities that might attack the data, and we hoped that they could mediate and guide the project through any problems which might arise. We recommend that other data stewards take similar steps to prepare for the unexpected, both in the mass collaborative setting and in the more common setting of providing data for use by individual researchers.

\section{Third-party guidance}
\label{guidance}

It is easy for those coordinating a project to fall into groupthink and ignore potential problems out of a common interest in the success of the project. To avoid this pitfall, we worked under the guidance of third parties. First, the Fragile Families Study and the Fragile Families Challenge were undertaken with the oversight of the Institutional Review Board at Princeton University.\footnote{Components of the Fragile Families Study were also reviewed by the Institutional Review Boards of partner institutions (i.e., Columbia University and Westat, the data collection contractor).}

We created an additional community for third-party guidance by assembling a Board of Advisors.  The Board included professors of sociology, education, and social work who had each devoted much of their careers to studying disadvantaged families, a journalism professor, a machine learning researcher, and a computational social scientist, some of whom are authors of this paper.  During our privacy and ethics audit we sent the Board weekly updates about our progress.  These weekly updates were also shared with staff working on the Fragile Families Study, as well as other stakeholders. We found this process of weekly updates incredibly valuable both to sharpen our thinking and to ensure that all stakeholders were involved. 

Finally, we also sought informal advice from a wide variety of people not involved in the Challenge in any way.  These outsiders included a philosophy professor, a member of the military with experience planning high-risk operations, a lawyer with experience dealing with health records, and a public-interest lawyer who provides direct services to children in foster care.  We found that these uninvolved third-parties often provided an interesting perspective, and we would encourage other data stewards to have broad discussions if possible.  Overall, we believe that third-party guidance helped improve our process.

\section{Ethical framework}
\label{ethics}

During our process we had to make many different decisions, and these decisions were guided by the principles described in the Belmont Report \citep{belmontReport}, a foundational document in social science research ethics.  More specifically, the Belmont Report emphasizes three principles that should guide research ethics: respect for persons, beneficence, and justice.

\subsection{Respect for persons}

The principle of respect for persons means that researchers should respect each participant's autonomy to choose whether to participate after being informed about the nature of the research.  This principle is often operationalized as the process of informed consent.  In this case, when parents and children agreed to participate in the Fragile Families Study, they understood that the data they provided would be used for scientific research.  Therefore, when designing and conducting the Challenge we remained within the context of scientific research as much as possible in order to honor the agreement between survey respondents and researchers.  For example, when structuring the Challenge we decided not to offer prize money because that is not generally consistent with the context of scientific research.  Our thinking about preserving context was heavily influenced by Nissenbaum's (\citeyear{nissenbaum2009}) work on the importance of ``contextual integrity.''

\subsection{Beneficence}

The second key principle discussed in the Belmont Report is beneficence: the potential harm of the study must be weighed against the potential benefits.  Further, reasonable steps should be taken to maximize benefits and minimize harms---both probability and magnitude.  These ideas were central to our process of threat modeling and threat mitigation.  For example, these ideas affected our changes to the data (see Section \ref{modifications}). We hoped to reduce the probability of harm by redacting or adding noise to variables that we expected might aid re-identification, and we hoped to reduce the magnitude of harm should re-identification occur by redacting some information (e.g., illegal behaviors). However, we made these decisions while also weighing the potential benefits of including variables that might help contribute to scientific knowledge. For instance, we did not redact information about child behavior problems because this information might be an especially important predictor of adolescent outcomes. The principle of beneficence thus affected not only our decision to release the data but also our decision about which variables ought to be released.  

In addition to risks and benefits for participants in the Fragile Families Study, we also considered possible broader social impacts of our work \citep{zook2017}.  Predictive models similar to those built during the Challenge are increasingly being used for high-stakes decisions---such as in criminal justice \citep{berk2017} and child protective services \citep{chouldechova2018}.  Although these predictive models 
can improve social welfare \citep{kleinberg2015,kleinberg2017}, they can also discriminate against protected groups \citep{barocas2016} and magnify social inequality \citep{eubanks2018}.  Therefore, we weighed the possibility that the knowledge created in the Challenge could be used inappropriately against the possibility that it could be used to help policymakers who are seeking to understand the strengths and weaknesses of predictive modeling.  Because of the type of data that we used and the outcomes we selected, we believe that the risk of unintended secondary use is low and that it is outweighed by the possible social benefits of the research.  Because the risks of the Challenge---both to survey participants and to society---were very difficult to quantify, we found that existing ethical, philosophical, and legal debates about the precautionary principle \citep{oriordan1994,sunstein2003,sunstein2005,narayanan2016} and dual use research \citep{national2004,selgelid2009} helped guide our thinking.

\subsection{Justice}

The final principle in the Belmont report is justice: risks and benefits should flow to similar populations. Unfortunately, many failures of scientific research ethics have involved disadvantaged or vulnerable populations (for examples see \citealt{emanuel2008}). Informed by this history, social scientists today recognize a special obligation to make ethical decisions about research involving vulnerable populations. The fact that some participants in the Fragile Families Study are disadvantaged or children (or both) caused us some concern.  However, the nature of the Challenge meant that the population most likely to benefit from the scientific knowledge produced by the Challenge was the very population from which participants were drawn.  In other words, by conducting research on a sample of disadvantaged urban families, we can generate knowledge that might help improve the life chances of future families in similar positions.  Thus, our approach to the principle of justice is heavily influenced by the argument that no group should be excluded from the potential to benefit from research \citep{mastroianni2001}.

\section{Decision}
\label{decision}

Ultimately, after all the discussing, designing, and debating, we faced a decision: whether to go forward with the Challenge or not.  When forced to make a go/no-go decision, we conducted a comprehensive re-assessment.  We believed that the project was consistent with existing ethical rules and ethical principles governing social science research.  This rules-based decision was made by our Institutional Review Board, and the principles-based decision was made by us and by members of our Board of Advisors.  We also believed that after our threat mitigation the risks were low in an absolute sense and in appropriate balance with the potential for scientific and societal benefit.

Before a full scale launch, however, we conducted a pilot test in an attempt to discover any errors in our threat modeling, threat mitigation, or ethical thinking. We conducted the pilot test by deploying the Challenge as a project in an undergraduate machine learning class at Princeton University. In doing so, we faced a difficulty inherent in pilot testing: being realistic while also being safe. By conducting the test in a class taught by a trusted professor, we erred on the side of safety. In our case, the pilot test turned out to be useful; it helped us discover one variable in the Challenge data file that was both confusing to participants and increased the risk of re-identification. The pilot also gave us a chance to test our screening process, which we modified for the full launch to gather more information and facilitate the process for both participants and organizers. The full launch of the Challenge was much smoother because we started in a controlled setting, and we highly recommend that others conduct a similar pilot test.

\section{Conclusion}
\label{conclusion}

This case study describes the privacy and ethics audit that we conducted as part of the Fragile Families Challenge.  Our process was certainly not perfect, and we hope that by being open about it, others can improve on what we did.\footnote{For more on a process-based approach to data access, see \citet{rubinstein2016}.}  We want to emphasize that other data stewards may reasonably come to different decisions about how to strike an appropriate balance in their own situation.  Despite differences between situations, however, we believe that the key elements of our approach---threat modeling, threat mitigation, and third-party guidance, all within a specific ethical framework---may be useful to other data stewards, whether they reside in universities, companies, or governments.

Those who might wish to follow or build on our example will undoubtedly wonder about the costs of doing so. The approach we followed was time-consuming.  We spent about three months preparing to launch the Challenge, and the privacy and ethics audit was the most time-consuming part (see Figure \ref{fig:timeline}).  The parts that took the most time were assembling a team with appropriate expertise (including for the response plan), attacking our data, debating how to balance the various trade-offs, and keeping in contact with stakeholders.  Attacking the data, by itself, involved about 1.5 months of consistent work for a skilled graduate student.  Beyond the amount of time involved, this process was also stressful and emotionally taxing; some of us found it difficult to spend so much time imagining all the worst-case scenarios of re-identification. 

Although our privacy and ethics audit was time-consuming, we feel that the effort was worthwhile.  In particular, this process enabled us to run the Challenge, which has already started to achieve some of its scientific objectives, as illustrated by the special issue in which this paper will be published.  

Finally, we hope that this case study illustrates that there is an important middle ground between calls for no data sharing and complete data sharing.  Everyone will benefit if scientists, companies, and governments can continue to develop the technical, legal, ethical, and social infrastructure to enable safe and responsible data access.

\newpage
\begin{singlespacing}
\bibliography{PrivAudit} 
\end{singlespacing}

\newpage
\section{Appendix}
This appendix includes reproductions of the web forms participants completed to (a) apply to participate and (b) agree to a set of terms and conditions of participation.

\begin{enumerate}
\item[a)] We used the application to assess each applicant's ability to contribute and motivations, and to gain advance notice of potential threats with whom to follow up.
\item[b)] We used the terms and conditions to reinforce norms of scientific behavior and emphasize the importance of respecting the data. The terms and conditions also include a set of items about how we would use the submissions of Challenge participants.
\end{enumerate}

\begin{center}
\includegraphics[page = 1, height = \textheight]{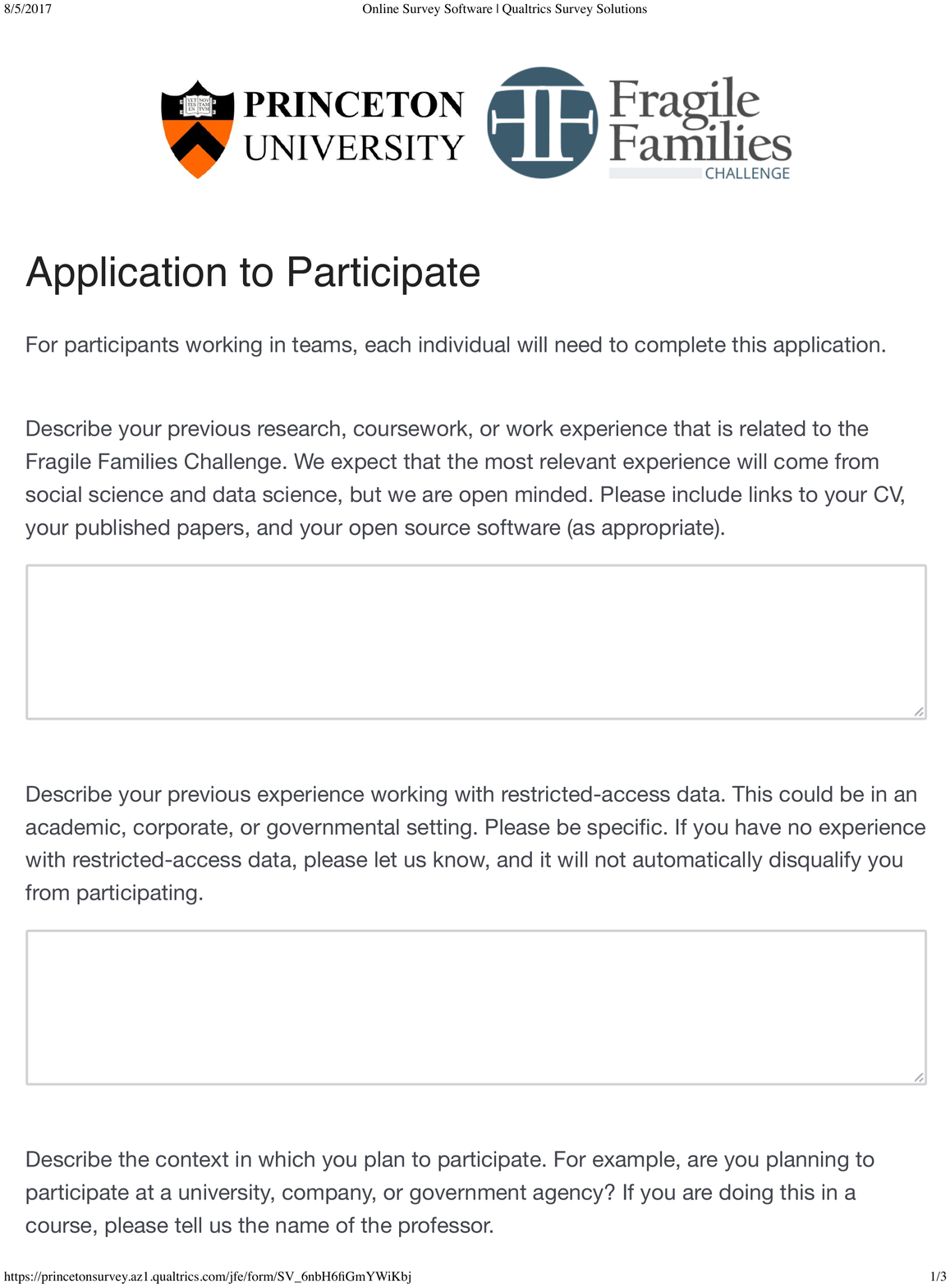}
\newpage
\includegraphics[page = 2, height = \textheight]{ApplicationA.pdf}
\newpage
\includegraphics[page = 1, height = \textheight]{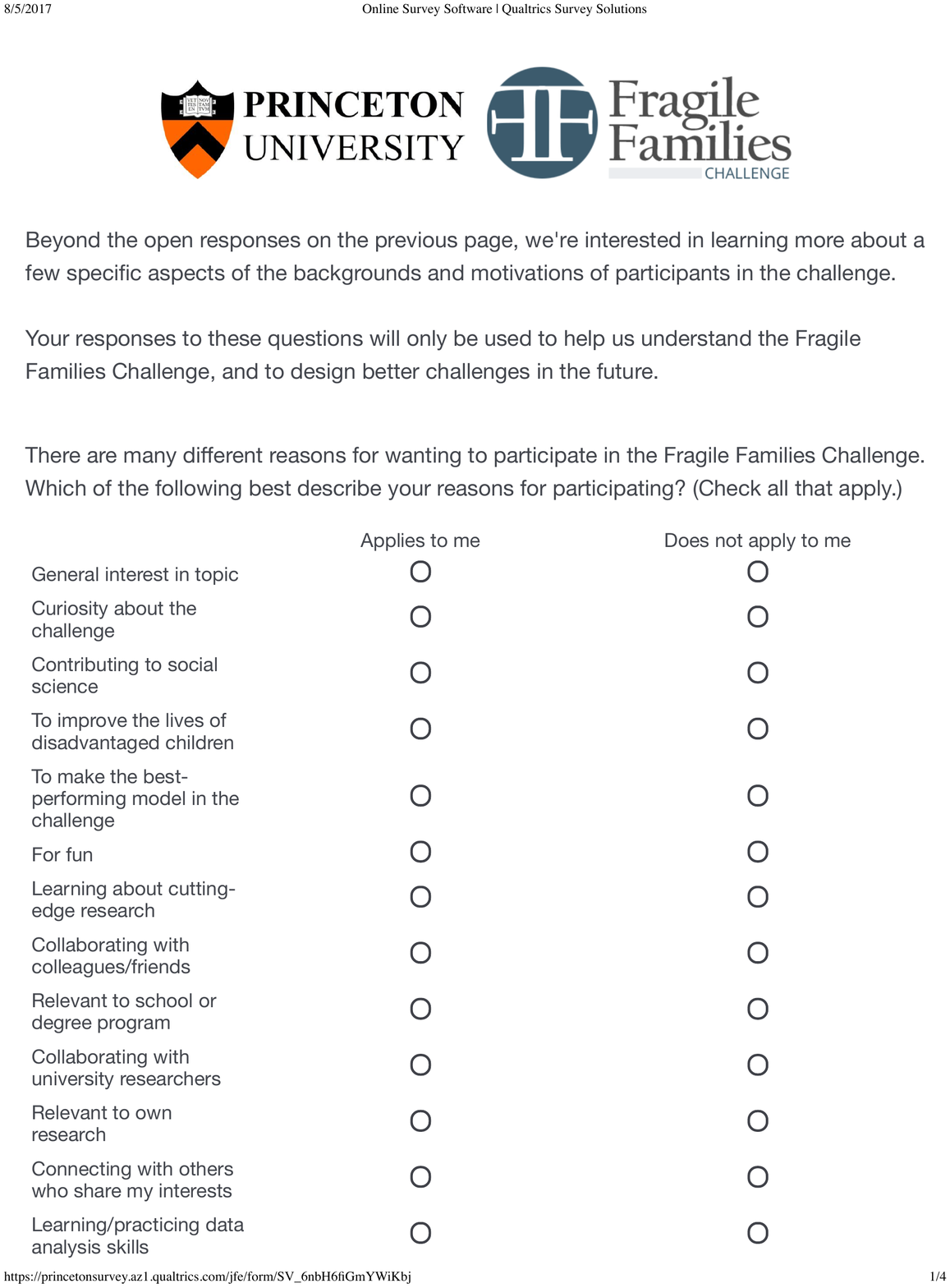}
\newpage
\includegraphics[page = 2, height = \textheight]{ApplicationB.pdf}
\newpage
\includegraphics[page = 3, height = \textheight]{ApplicationB.pdf}
\newpage
\includegraphics[page = 1, height = \textheight]{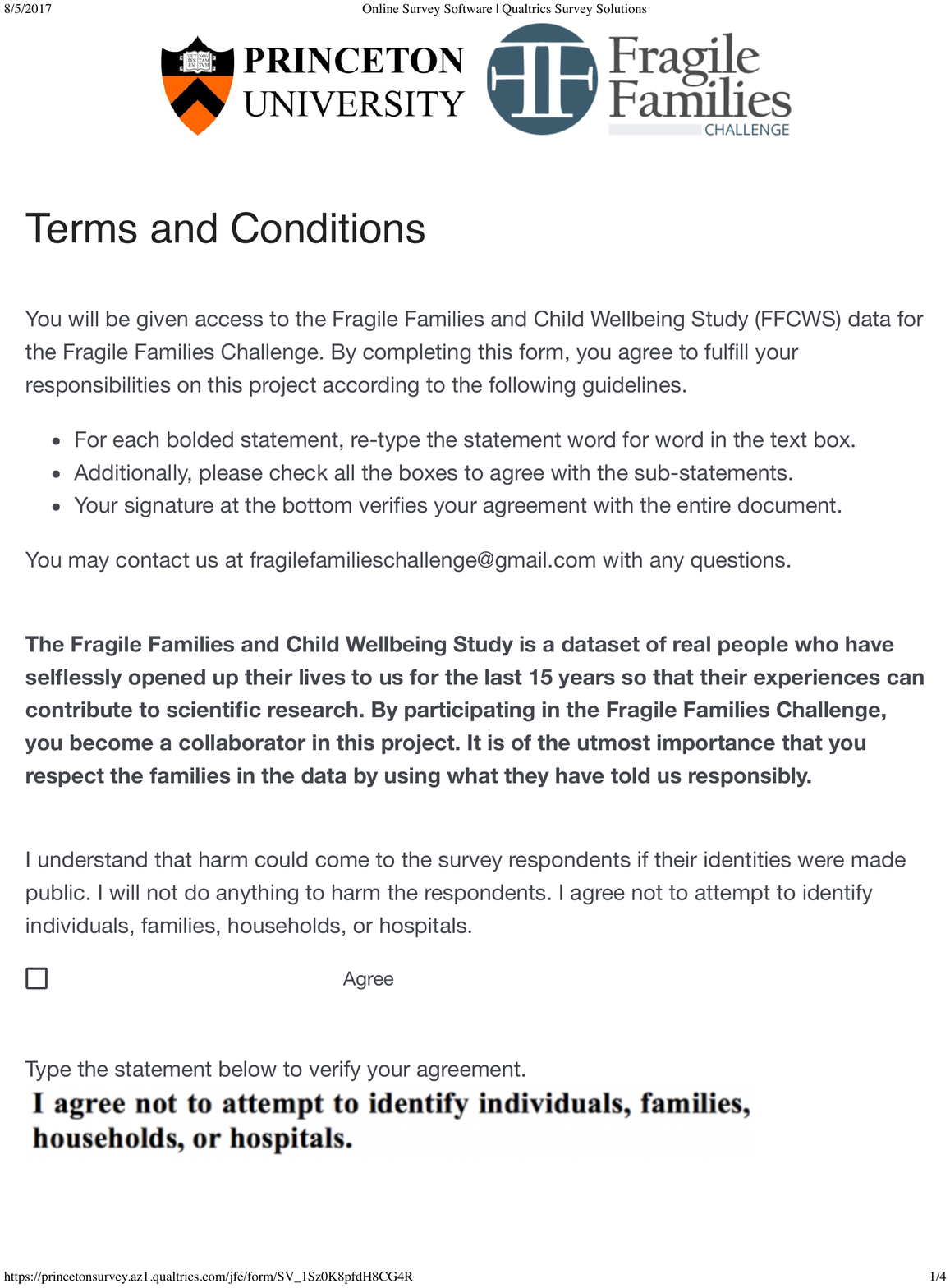}
\newpage
\includegraphics[page = 2, height = \textheight]{TermsAndConditions.pdf}
\newpage
\includegraphics[page = 3, height = \textheight]{TermsAndConditions.pdf}
\newpage
\includegraphics[page = 1, height = \textheight]{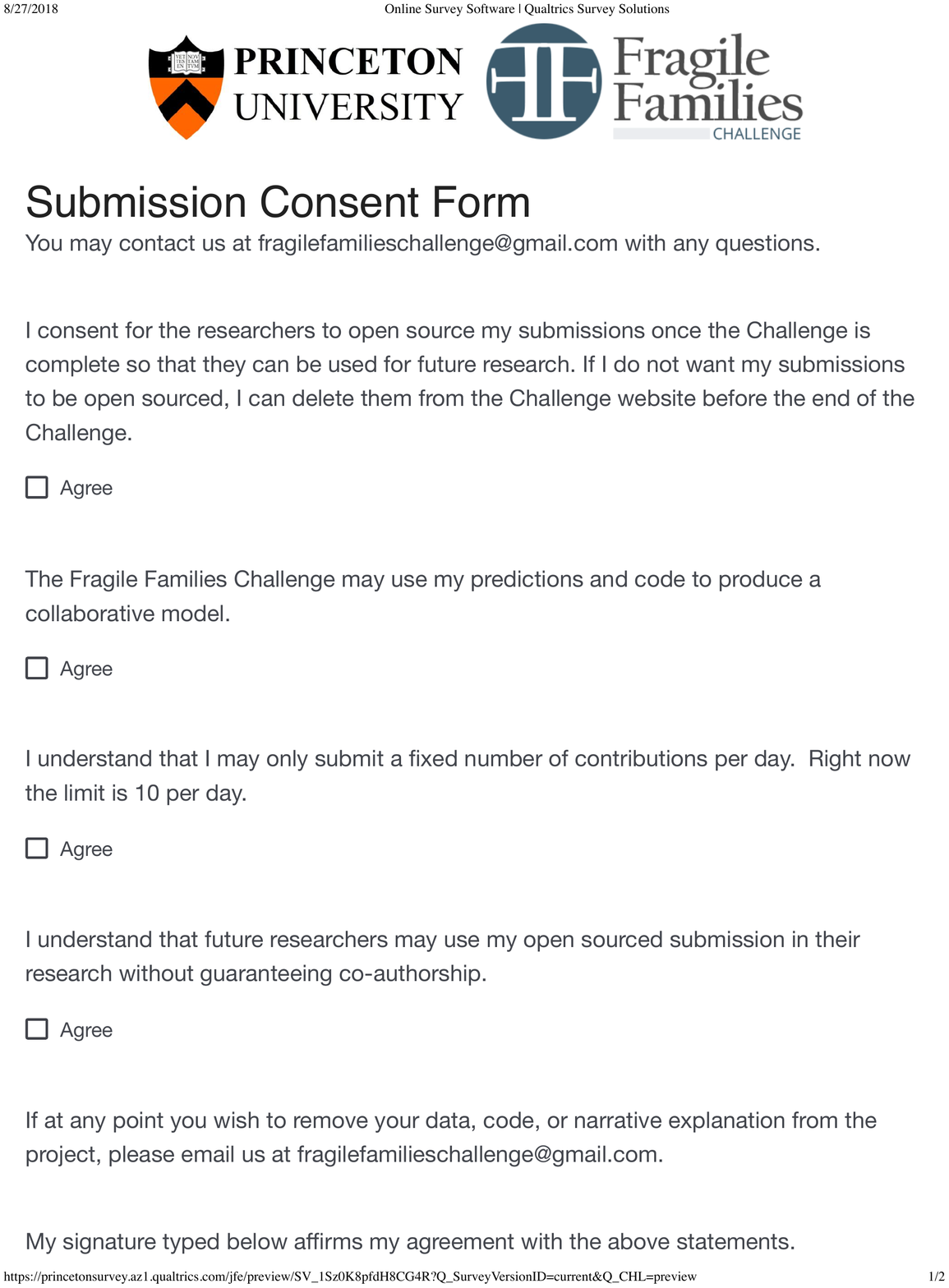}
\newpage
\includegraphics[page = 2, height = \textheight]{submission_consent_form.pdf}

\end{center}

\end{document}